\documentclass[%
 reprint,
 amsmath,amssymb,
 aps,
 prb,
floatfix,
showpacs,
showkeys
]{revtex4-1}

\usepackage{graphicx}
\usepackage{dcolumn}
\usepackage{bm}

\begin{document}

\preprint{MAN_MSU_CUP-2017-1}

\title{Quantum Dragon Solutions for Electron Transport through Nanostructures based on Rectangular Graphs}%

\author{G.\ Inkoom}
\author{M.A.\ Novotny}%
\altaffiliation[Also at ]{Faculty of Mathematics and Physics, Charles University, Ke Karlovu 5, CZ-121 16 Praha 2, Czech Republic}%
\email{man40@msstate.edu}
\affiliation{%
 Department of Physics and Astronomy\\
 Mississippi State University\\
 Mississippi State, MS 39762-5167
 USA 
}%

\date{\today}

\begin{abstract}
Electron transport through nanodevices of atoms in a single-layer rectangular arrangement 
with free (open) boundary conditions parallel to the direction of the current flow is studied
within the single-band tight binding model.  
The Landauer formula gives the electrical conductance to be a function of the electron 
transmission probability, ${\cal T}(E)$, as a function of the energy $E$ of the incoming electron.  
A quantum dragon nanodevice is one which has a perfectly conducting channel, 
namely ${\cal T}(E)=1$ for all energies which are transmitted 
by the external leads even though there may be arbitrarily strong electron scattering.  
The rectangular single-layer systems are shown to be able to be quantum dragon devices, 
both for uniform leads and for dimerized leads.  
The quantum dragon condition requires appropriate lead-device connections and 
correlated randomness in the device.  
\end{abstract}

\maketitle


\section{\label{Sec:01} Introduction}

The confluence of the approaching end of Moore's law for electronics \cite{WALD2016} 
and the increasing ability of experimentally being able to manipulate, fabricate, and measure 
at the nanoscale level \cite{TAKA2010} 
is the reason rapid progress in nanoelectronics is being made \cite{PUER2017}.  
Of particular interest is the electron transmission properties of 
nanodevices \cite{DATTA1995,FerryGoodnick1997,TODO2002,DATTA2005}, 
including electron transmission in molecular electronics \cite{ZIMB2011,CUEV2017}.  

Due to the quantum mechanics underlying nanodevices, properties of coherent 
electron transport can be very different from those expected at the macroscopic scale.  
As shown by Landauer \cite{LAND57}, of central importance to nanodevices is the 
electron transmission, ${\cal T}$, of the nanodevice when it is connected to leads attached 
to a source and a sink of electrons.  As in macroscopic systems, one desires the 
electrical conductance $G$ (the inverse of the electrical resistance) in an Ohm's law 
relationship $I=G V$ where $I$ is the electrical current flowing through the device and 
$V$ is the applied electrical voltage difference.  At low temperatures the Landauer 
formula gives the electrical conductance \cite{FerryGoodnick1997,BAGW1989}
\begin{equation}
\label{Eq:LandauerG}
G = \left\{
\begin{array}{lcl}
G_0 \> {\cal T}(E_F) & \qquad & {\rm two\> probe} \\
G_0 \> \frac{{\cal T}(E_F)}{1-{\cal T}(E_F)} & \qquad & {\rm four\> probe} \\
\end{array}
\right.
\end{equation}
for two probe or four probe measurements.  Here $G_0=\frac{2 e^2}{h}$ is the 
quantum of conductance, with $e$ the charge of the electron, $h$ Planck's constant, 
and the factor of two is due to the spin of the electron.  
The transmission is a function of the energy $E$ of an incoming electron, 
and in Eq.~(\ref{Eq:LandauerG}) the transmission at the Fermi energy $E_F$ enters.  
The power of the shot noise of the nanodevice is \cite{LESO1989,BUTT1990,KUMA1996,OUIS2013}
\begin{equation}
\label{Eq:ShotNoise}
P \> = \> \frac{4 e^3}{h} \>{\cal T}(E_F) \> \left[1-{\cal T}(E_F)\right] \> V
\end{equation}
which is zero if ${\cal T}(E_F)$$=$$1$.  

In this investigation we are interested in a perfectly conducting channel, namely 
where ${\cal T}(E)=1$ for all energies $E$ which can propagate through long 
leads.  There are different possibilities for the existence of perfectly conducting channels 
in coherent electron propagation.  These possibilities include:
\begin{itemize}
\item {\bf Ballistic propagation\/}.  
\subitem If there is no scattering in the device, 
electrons propagate ballistically, and ${\cal T}(E)=1$.  For ballistic propagation, 
the two different behaviors for $G$ in Eq.~(\ref{Eq:LandauerG}) have been observed 
experimentally in the same sample of a very pure semiconductor \cite{DEPI2001}.  
\item {\bf Long-range randomness\/}.  
\subitem For example, in zigzag carbon nanoribbons when there is random long-range scattering, 
the average $\langle G\rangle\rightarrow G_0$ in two probe measurements as the 
length $L$ of the device increases \cite{WAKA2007}.
\item {\bf Surface states in topological insulators\/}.  
\subitem  In topological insulators, surface states can lead to perfectly 
conducting channels which are protected against scattering due to 
disorder \cite{MATS2015}.  
\item {\bf Quantum dragons\/}.  
\subitem In 2014, one of the authors discovered a 
large class of nanodevices \cite{MANdragon2014}
which can have arbitrarily strong scattering, 
but because the scattering is correlated they can have ${\cal T}(E)$$=$$1$ for any $L$.  
These nanodevices, with strong scattering but with 
a perfectly conducting channel, were named \lq quantum dragons' in ref.~\onlinecite{MANdragon2014}.  
\end{itemize}
Sometimes nanodevices with a perfectly conducting channel have been said to 
be metallic or ballistic, due to the electrical conductance given by Eq.~(\ref{Eq:LandauerG}).  Examples 
are armchair single walled carbon nanotubes \cite{OUYA2001,KONG2001} and graphene nanoribbons 
\cite{BARI2014,CELI2016}.  TEM and SEM investigations of gold point contacts 
have also been performed \cite{Erts2000}.

Single layer thick carbon nanotubes and graphene nanoribbons have been fabricated, 
and their intriguing properties studied \cite{BARI2014,CELI2016}.  These systems are all based 
on an underlying hexagonal lattice.  Single-walled carbon nanotubes in the armchair arrangement 
have been experimentally shown to have metallic behavior \cite{OUYA2001}, 
often in the literature said to exhibit ballistic electron propagation \cite{WHIT1998}.  
However, one needs to be careful to 
remember ${\cal T}(E)=1$ only if an armchair single-walled carbon nanotube is connected 
to appropriate leads in the correct fashion \cite{OUYA2001,MANdragon2014}.  

The name quantum dragons denotes such devices may be formed by joining different 
types of nanodevices, their length is typically longer than any other dimension, and 
they are invisible to electrons which propagate in the leads.  
The quantum dragon nanodevices published in ref.~\onlinecite{MANdragon2014} all had cylindrical symmetry.   
In this paper we investigate quantum dragon nanodevices without cylindrical symmetry.  
In particular, we investigate quantum dragons in single-layer thick nanodevices based 
on an underlying rectangular graph, with open (free) boundary conditions perpendicular 
to the direction of the current flow.  

Recently, some single layer thick materials based on rectangular lattices have been 
synthesized.  
One example is free-standing single-atom-thick iron membranes \cite{ZHAO2014}, 
and another is copper oxide monolayers \cite{YIN2016,KANO2017}.
Other examples are 2D materials and van der Waal heterostructures have recently 
been reviewed \cite{NOVO2016}.  
These experimental systems, and the possibility of many more 2D systems based on 
rectangular lattices, provide the impetus to study whether nanodevices based on 
underlying rectangular graphs can have ${\cal T}(E)$$=$$1$, and in particular whether 
such systems can be quantum dragons.  

The paper is organized as follows.
In Sec.~\ref{Sec:02}, the method of calculating ${\cal T}(E)$ for non-dimerized 
devices and leads is presented.  The method used is the matrix method \cite{DCA2000}.  
Sec.~\ref{Sec:03} gives the method to obtain quantum dragon solutions via 
the exact mapping and tuning process for systems based on rectangular graphs.  
Sec.~\ref{Sec:04} contains example numerical calculations of non-dimerized quantum dragons, 
thereby better illustrating the concept.  
Sec.~\ref{Sec:05} has our conclusions and further discussion.  
The main text is supplemented with a number of appendices.  
A detailed discussion of the example devices shown is presented in App.~A 
for devices without disorder and in App.~B 
for devices with disorder.  
The matrix method to calculate ${\cal T}(E)$ for dimerized leads 
is derived in App.~C. 
Quantum dragon solutions for nanodevices based on rectangular graphs with dimerized leads is
presented in App.~D. 
The relationship between the matrix method used in this article and the commonly 
used Green's function method to calculate ${\cal T}(E)$ is given in App.~E. 
Appendix~F 
shows in mathematical detail how quantum dragon solutions arise in the case of 
two slices in the nanodevice.

\begin{center}
\begin{figure}[b]
\vspace{0.05 cm}
\includegraphics[width=0.40\textwidth]{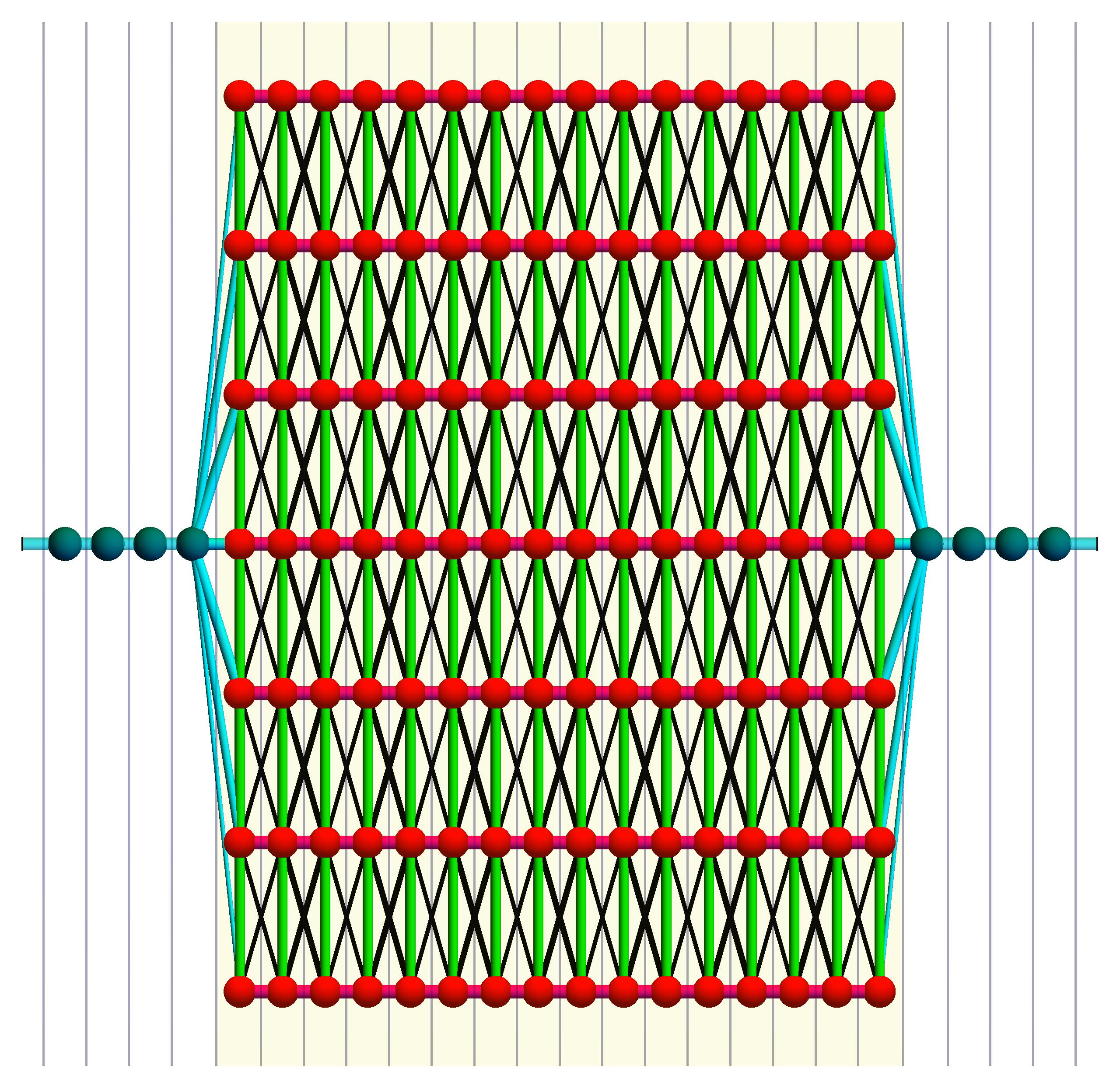}
\caption{
\label{Fig:SQdragon:Fig01}
(Color online.) 
An example of leads connected to a planar rectangular nanodevice
with $\ell=16$ and $m=7$.  
Leads and device are both uniform and without disorder.  
See Appendix~A.1 
for a complete description.  
}
\end{figure}
\end{center}

\section{\label{Sec:02} Transmission ${\cal T}(E)$ Calculations}

The electron transmission is found from the solution of the time-independent 
Schr{\"o}dinger equation for the nanodevice and leads.  Although solutions 
using density functional theory, 
as in the review \cite{ZIMB2011}, are possible, this imposes a limit on the number of 
atoms which can be studied in the nanodevice and furthermore limits one to 
numerical investigations.  Therefore, here we study the nanodevice and leads within 
the single-band tight binding (TB) model.  In the TB model the important quantities are 
the on~site energy (at the location of every atom), $\epsilon$, and the hopping 
between atoms, denoted in this article by either $t$ or $s$.  
The hopping terms come from overlaps of electron 
wavefunctions between atoms located near one another, so in the device and the leads 
we limit ourselves to nearest-neighbor (nn) and next-nearest-neighbor (nnn) hopping terms.  
In most cases the hopping terms are negative, so we put in the negative sign \lq by hand' 
and let $t$ or $s$ stand for the magnitude of the hopping.  
Four advantages of the TB model for electron transport calculations are detailed 
in ref.~\onlinecite{TODO2002}.
We only analyze the two terminal measurement setup.
Within the TB model, the matrix equation to solve is 
\begin{equation}
\label{EQ:S2:Sch}
\left({\cal H} - E {\bf I}_\infty \right) {\vec \Psi} \> = \> {\vec 0} 
\end{equation}
where ${\bf I}_\infty$ is the infinite identity matrix, 
${\cal H}$ is the infinite matrix for the two semi-infinite leads and the device, 
and $E$ is the energy of the incoming electrons.

In the text in the main article, we concentrate on uniform (not dimerized) 
leads attached to a rectangular device, as in Fig.~\ref{Fig:SQdragon:Fig01}.  
The case where the leads are dimerized has the matrix method solution 
derived in App.~C 
and analyzed in App.~D. 
We have freedom in choosing our 
zero of energy, so we choose the on site energy of the lead atoms to be zero.  
The hopping strength between lead atoms we take to be $t_{lead}$.  In many cases 
theorists take $t_{lead}$ to be the unit of energy and set it to 
unity \cite{MANdragon2014,DCA2000,BOET2011,NOVO2014}, but we 
will keep $t_{lead}$ throughout in order to make better connection with the 
dimerized leads in the appendices.  We assume our nanodevice has an underlying 
rectangular graph, as in Fig.~\ref{Fig:SQdragon:Fig01}. 
Every slice (every column) of the nanodevice has $m$ atoms, and there are $\ell$ 
slices.  Within column $j$, every atom has the same on site energy $\epsilon_j$ 
and the same nn hopping $t_j$.  
Between columns $j$ and $j+1$ there can be nn hopping of strength 
$s_{nn,j}$ and nnn hopping of strength $s_{nnn,j}$.  
The device can be considered to be a planar rectangular 
array of atoms when there is no disorder, as in Fig.~\ref{Fig:SQdragon:Fig01}, 
when all TB parameters are the same, 
that is $\epsilon_j$$=$$\epsilon$, $t_j$$=$$t$, $s_{nn,j}$$=$$s_{nn}$, and $s_{nnn,j}$$=$$s_{nnn}$.  

\begin{center}
\begin{figure*}[tbh]
\vspace{0.05 cm}
\includegraphics[width=0.90\textwidth]{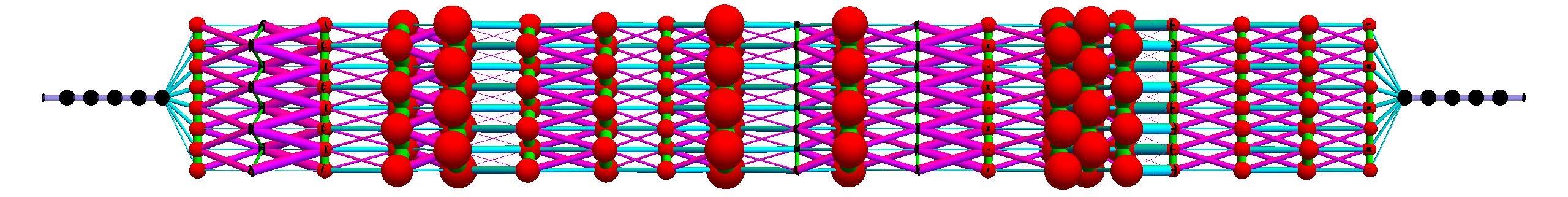}
\\
~~~~
\\
\includegraphics[width=0.90\textwidth]{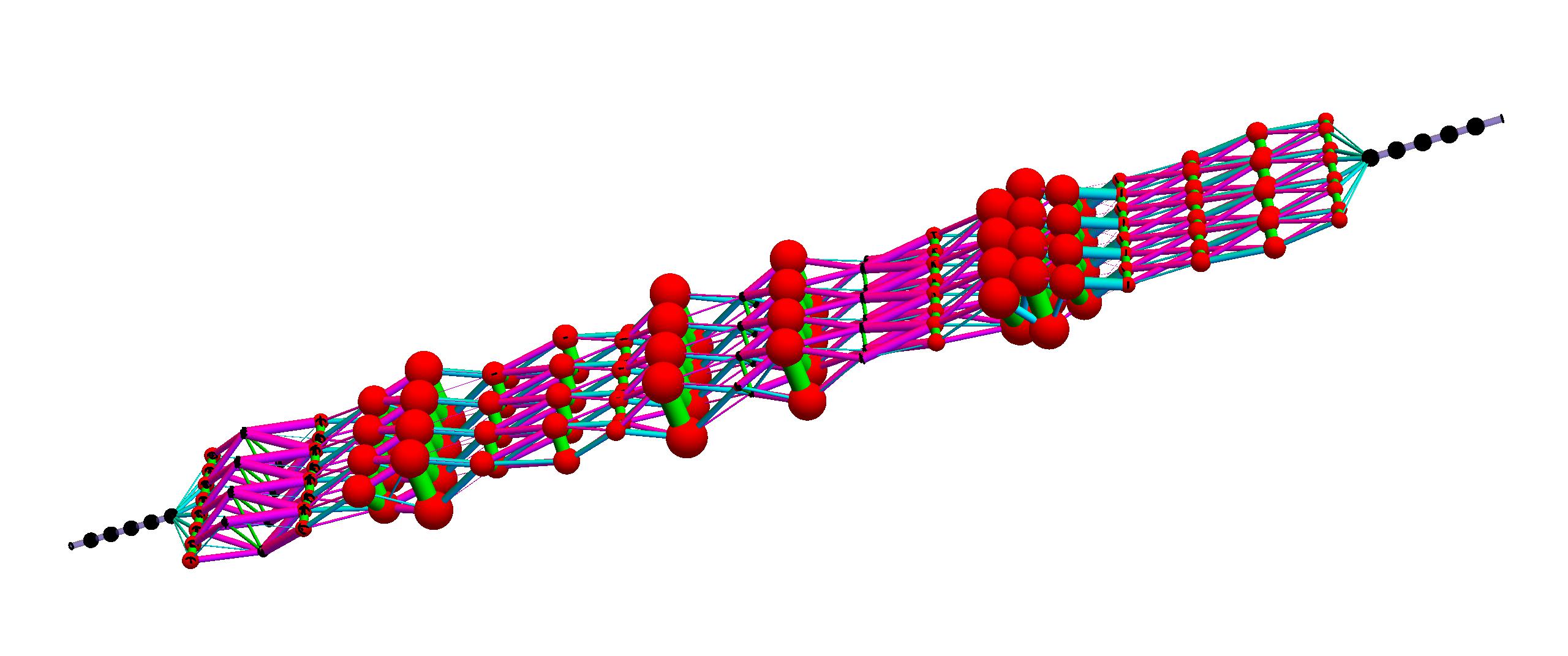}
\caption{
\label{Fig:SQdragon:Fig02}
(Color online.) 
An example of uniform leads connected to a disordered, rectangular device 
with $m$$=$$8$ and $\ell$$=$$20$.  
The same device is shown in the top (top view) and 
bottom (oblique view) of the figure.  
See Appendix~B 
for a complete description.  
}
\end{figure*}
\end{center}

When there is disorder in the TB parameters $\epsilon_j$, $t_j$, $s_{nn,j}$ 
and $s_{nnn,j}$, the underlying graph is still rectangular but the 
nanodevice need not remain planar, as in Fig.~\ref{Fig:SQdragon:Fig02}.  

We use the matrix method \cite{DCA2000} to calculate the transmission, ${\cal T}(E)$, because 
the mapping and tuning method to find quantum dragons takes advantage of 
the matrix structure.  App.~E 
gives the relationship between the commonly used 
Green's function method 
\cite{DATTA1995,FerryGoodnick1997,TODO2002,DATTA2005,ZIMB2011,CUEV2017,TRIO2016} 
and the matrix method.  
For dimerized leads the method is derived in App.~C, 
while for uniform leads the matrix method was put forward in 2000 \cite{DCA2000}.   

Unless otherwise explicitly stated or indicated by subscripts, the dimension of 
all vectors is $m$ and all matrices are of size $m$$\times$$m$.  The transmission 
is calculated by ${\cal T}$$=$$\left|t_T\right|^2$ where $t_T$ is calculated by the 
inverse of a $(\ell m$$+$$2)$$\times$$(\ell m$$+$$2)$ matrix ${\bf M}_\ell$ which has the form 
(written for $\ell$$=$$4$) 
\begin{widetext}
\begin{equation}
\label{EQ:S2:Tlarge}
{\bf M}_4 
\left(\begin{array}{c}
1 + r  \\
{\vec \psi}_1 \\
{\vec \psi}_2 \\
{\vec \psi}_3 \\
{\vec \psi}_4 \\
t_T \\
\end{array}\right)
\> = \> 
\left(\begin{array}{cccccc}
\! \xi & {\vec w}^\dagger & {\vec 0}^\dagger & {\vec 0}^\dagger & {\vec 0}^\dagger & 0 \\
{\vec w} & {\bf F}_1 & {\bf B}_{12} & {\bf 0} & {\bf 0} & {\vec 0} \\
{\vec 0} & {\bf B}_{12}^\dagger & {\bf F}_2 &  {\bf B}_{23} & {\bf 0} & {\vec 0} \\
{\vec 0} & {\bf 0} & {\bf B}_{23}^\dagger & {\bf F}_3 &  {\bf B}_{34} & {\vec 0} \\
{\vec 0} & {\bf 0} & {\bf 0} & {\bf B}_{34}^\dagger & {\bf F}_4 & {\vec u} \\
0 & {\vec 0}^\dagger & {\vec 0}^\dagger & {\vec 0}^\dagger & {\vec u}^\dagger & \! \xi \! \\
\end{array}\right)
\!\!
\left(\begin{array}{c}
1 + r \\
{\vec \psi}_1 \\
{\vec \psi}_2 \\
{\vec \psi}_3 \\
{\vec \psi}_4 \\
t_T \\
\end{array}\right)
= 
\left(\begin{array}{c}
\!\! \Lambda \!\! \\
\!\! {\vec 0} \!\! \\
\!\! {\vec 0} \!\! \\
\!\!{\vec 0} \!\! \\
\!\! {\vec 0} \!\! \\
\!\! 0 \!\! \\
\end{array}\right)
\end{equation}
\end{widetext}
where the vector ${\vec w}$ (${\vec u}$) contains the TB hopping terms 
between the incoming left (outgoing right) lead and the atoms in the 
first (last) slice of the device.  
The intra-slice TB terms, here $\epsilon_j$ and $t_j$, are in the 
matrices ${\bf F}_j={\bf A}_j-E {\bf I}$ with ${\bf I}$ the $m$$\times$$m$ 
identity matrix.  
Note the energy $E$ is only present in the diagonal elements of ${\bf M}_\ell$.  
The TB intra-slice matrix is 
\begin{equation}
\label{EQ:S3:DefineA}
{\bf A}_j  \> = \> \epsilon_j {\bf I} - t_j {\bf Q} 
\quad {\rm for} \> j=1,2,\cdots,\ell
\end{equation}
and the inter-slice TB terms are 
\begin{equation}
\label{EQ:S3:DefineB}
{\bf B}_{j,j+1} \> = \> -s_{nn,j} {\bf I} - s_{nnn,j} {\bf Q} 
\>.
\end{equation}
The matrix ${\bf Q}$ is defined by 
\begin{equation}
\label{EQ:S3:DefineQ}
{\bf Q} \> = \> 
\left(\begin{array}{ccccccc}
0 & 1 & 0 & \cdots & 0 & 0 & 0 \\
1 & 0 & 1 &        & 0 & 0 & 0 \\
0 & 1 & 0 & \cdots & 0 & 0 & 0 \\
\vdots &  & \vdots & \ddots & \vdots & & \vdots \\
0 & 0 & 0 & \cdots & 0 & 1 & 0 \\
0 & 0 & 0 &        & 1 & 0 & 1 \\
0 & 0 & 0 & \cdots & 0 & 1 & 0 \\
\end{array}\right)
\end{equation}
and has this form because we study a rectangular lattice, 
allow only nn hopping within each slice, and allow only nn and nnn 
hopping between slices.  
The matrix ${\bf A}_j$ contains the part of the Hamiltonian which 
includes all TB parameters within slice $j$, while ${\bf B}_{j,j+1}$ 
is the part of the Hamiltonian containing the TB parameters which are 
the hopping terms between atoms in slices numbered $j$ and $j$$+$$1$.  
The wavevector $q$ of the electron in the leads is given by 
\begin{equation}
\label{Eq:cosq}
\cos(q) = -\frac{E}{2 \> t_{lead}}
\end{equation}
where the distance has been taken to be unity between lead atoms.  
Hence $\sin(q)=\pm\sqrt{4 t_{lead}^2 - E^2}/2 t_{lead}$.  
Furthermore, in Eq.~(\ref{EQ:S2:Tlarge}) 
$\Lambda=-2 i \sin(q)$ and $\xi=e^{- i q}$.
For propagating modes we require 
$-1\le \cos(q)\le 1$, which give propagating modes 
for energies $-2 t_{lead}\le E\le 2 t_{lead}$.

\section{\label{Sec:03} Quantum Dragon Solutions via Exact Mapping}

In ref.~\onlinecite{MANdragon2014}, an exact mapping between a nanodevice with $\ell m$ atoms 
and a 1D (one dimensional) nanodevice with $\ell$ atoms was shown to be sometimes possible.  
The mapping preserves ${\cal T}(E)$, and provides a method to find the 1D chain of 
length $\ell$ which has the same ${\cal T}(E)$ as the nanodevice with $\ell m$ atoms.  
For some TB parameters in the original nanodevice, the 1D chain mapped onto is identical 
to a segment of length $\ell$ of the leads.  Consequently, for nanodevices with these 
parameters, one has ${\cal T}(E)$$=$$1$ since the 1D mapped system is indistinguishable from a 
segment of the same length of the leads.  For uniform leads, as we study in the main article, 
\cite{MANdragon2014} gives the requirements for the exact mapping, as well as conditions 
for a quantum dragon.  
The matrix method for the solution of ${\cal T}(E)$ and the mapping and tuning for 
dimerized leads and a dimerized device is given in App.~C and App.~D.  

The important consideration for the existence of an exact mapping when all slices have 
$m$ atoms is that there exists a vector ${\vec x}_1$ which is simultaneously an 
eigenvector of all ${\bf A}_j$ and all ${\bf B}_{j,j+1}$.  
Note the ${\bf A}_j$ are Hermitian, while the ${\bf B}_{j,j+1}$ do not need 
to be Hermitian.  
Furthermore, the nanodevice must be connected to the uniform leads such that 
${\vec w}\propto {\vec x}_1$ and ${\vec u}\propto {\vec x}_1$.  
Of course for physical nanodevices, ${\vec x}_1$ must be composed of valid TB hopping parameters.  
In our case, the ${\bf A}_j$ and ${\bf B}_{j,j+1}$ are tridiagonal Toeplitz matrices.  
We restrict ourselves to the case of zero magnetic field, so all TB hopping terms are real.  
We also restrict nnn hopping between atoms in slice $j$ and $j$$+$$1$ to all be identical, 
so in our case the ${\bf B}_{j,j+1}$ matrices are also Hermitian.  
All our matrices have the form 
\begin{equation}
\label{EQ:S3:GeneralIQ}
{\bf C} \> =c_{I} {\bf I} + c_{Q} {\bf Q} 
\end{equation}
for some parameters $c_{I}$ and $c_{Q}$, and the matrix ${\bf Q}$ is 
given in Eq.~(\ref{EQ:S3:DefineQ}).
The matrix in Eq.~(\ref{EQ:S3:GeneralIQ}) has eigenvalues \cite{JAIN1979}
\begin{equation}
\label{EQ:S3:Ceigenvalues}
{\widehat\lambda}_{k} \> = \> c_{I} + 2 c_{Q} \cos\left(\frac{k \> \pi}{m+1}\right)
\>\>\>\>\>\>\> {\rm for} \>\> 
k=1,2,\cdots,m \\
\end{equation}
and the eigenvector associated with ${\widehat\lambda}_1$ is  
\begin{equation}
\label{EQ:S3:Aeigenvectors}
{\vec x}_1 \> = \> 
\sqrt{\frac{2}{m+1}} \> 
\left(\begin{array}{c}
\sin\left(\frac{\>\> \pi}{m+1}\right) \\
\sin\left(\frac{2\> \pi}{m+1}\right) \\
\vdots \\
\sin\left(\frac{m\> \pi}{m+1}\right) \\
\end{array}\right)
\end{equation}
where we note that ${\vec x}_1$ is independent of 
$c_I$ and $c_Q$ \cite{JAIN1979}.  Furthermore, all elements in 
${\vec x}_1$ have the same sign, which is expected by the 
Perron-Frobenius theorem for non-positive (or non-negative) matrices.  
Although all eigenvectors and eigenvalues for the matrix ${\bf C}$ are known, 
to obtain a quantum dragon one only needs the single eigenvector 
${\vec x}_1$ and its associated eigenvalue.  
Connecting the device based on a rectangular graph to the leads with 
${\vec w}={\vec u}=-{\vec x}_1$ and using the vector ${\vec x}_1$ for 
the mapping allows an exact mapping onto a 1D device with the same 
${\cal T}(E)$.  
This exact mapping is for any nanodevice based on a rectangular graph 
for any intra-slice TB parameters $\epsilon_j$ and $t_j$ and 
any inter-slice TB parameters $s_{nn,j}$ and $s_{nnn,j}$.  
The exact mapping exists whether the lattice can be viewed as planar 
when all TB parameters are independent of the slice index (as 
in Fig.~\ref{Fig:SQdragon:Fig01}), or ones which are most likely non-planar 
when the TB parameters depend on the slice index $j$, as in 
Fig.~\ref{Fig:SQdragon:Fig02}.  

Now that the exact mapping has been found, the only question to ask is 
what TB parameters give the mapped 1D system with the same TB parameters 
as the lead, namely on site energies $\epsilon_{lead}=0$ and hopping 
strengths $t_{lead}$.  These systems will be quantum dragons.  
The eigenvalues of the intra-slice matrices ${\bf A}_j$ of Eq.~(\ref{EQ:S3:DefineA})
associated with eigenvector 
${\vec x}_1$ have eigenvalues from Eq.~(\ref{EQ:S3:Ceigenvalues})  
\begin{equation}
\label{EQ:S3:Aeigenvalues}
\lambda_{j} \> = \> \epsilon_j - 2 t_j \cos\left(\frac{\pi}{m+1}\right)
\>\>\>\>\>\>\> {\rm for} \>\> 
j=1,2,\cdots,\ell 
\>.
\end{equation}
Similarly, from Eq.~(\ref{EQ:S3:Ceigenvalues}) and Eq.~(\ref{EQ:S3:DefineB}) 
the inter-slice matrices ${\bf B}_{j,j+1}$ have eigenvalues 
\begin{equation}
\label{EQ:S3:Beigenvalues}
\eta_{j} \> = \> -s_{nn,j} - 2 s_{nnn,j} \cos\left(\frac{\pi}{m+1}\right)
\end{equation}
for $j=1,2,\cdots,\ell\!-\!1$.  
The on site energies of the 1D mapped system are $\lambda_j$, and therefore 
a quantum dragon requires all $\lambda_j=0$, and hence from Eq.~(\ref{EQ:S3:Aeigenvalues})
\begin{equation}
\label{Eq:S4:Adragon}
\epsilon_j=2 t_j \cos\left(\frac{\pi}{m+1}\right) 
\qquad {\rm for} \>j=1,2,\cdots,\ell
\>.
\end{equation}
The mapped 1D system has hopping parameters between the mapped 1D device 
atoms equal to $\eta_j$, and therefore a quantum dragon requires all 
$\eta_j=-t_{lead}$, and therefore from Eq.~(\ref{EQ:S3:Beigenvalues})
\begin{equation}
\label{Eq:S4:Bdragon}
t_{lead} = s_{nn,j}+ 2 s_{nnn,j} \cos\left(\frac{\pi}{m+1}\right) 
\end{equation}
for all $j=1,2,\cdots \ell\!-\!1$.  
Furthermore, for a quantum dragon the connection vectors must be given by 
\begin{equation}
\label{Eq:S4:wudragon}
{\vec w}={\vec u}=-t_{lead} {\vec x}_1
\>.
\end{equation}
For every slice $j$ the intra-slice nn hopping $t_j$ may be any random value, provided 
for a quantum dragon 
the on site energy of slice $j$ satisfies Eq.~(\ref{Eq:S4:Adragon}).  
Similarly for the inter-slice hopping terms $s_{nn,j}$ and 
$s_{nnn,j}$ may be any random values provided they are correlated 
to satisfy Eq.~(\ref{Eq:S4:Bdragon}).  Therefore for every value of the index $j$, 
Eq.~(\ref{Eq:S4:Adragon}) 
and Eq.~(\ref{Eq:S4:Bdragon}) define what we mean by correlated randomness.  

\section{\label{Sec:04}Numerical calculation of uniform quantum dragons}
\begin{center}
\begin{figure}[tbh]
\vspace{0.05 cm}
\includegraphics[width=0.40\textwidth]{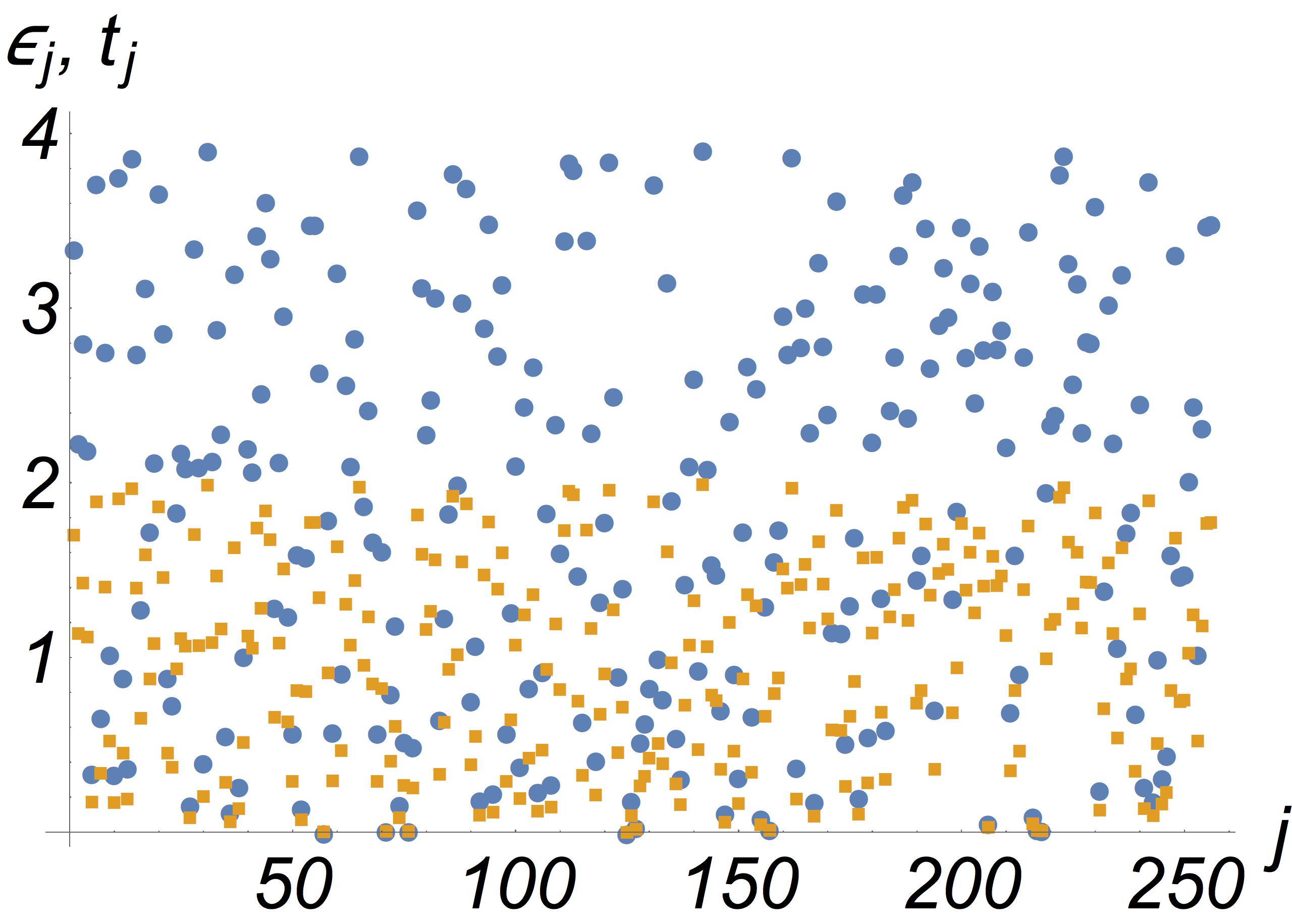}
\\
~~~~
\\
\includegraphics[width=0.40\textwidth]{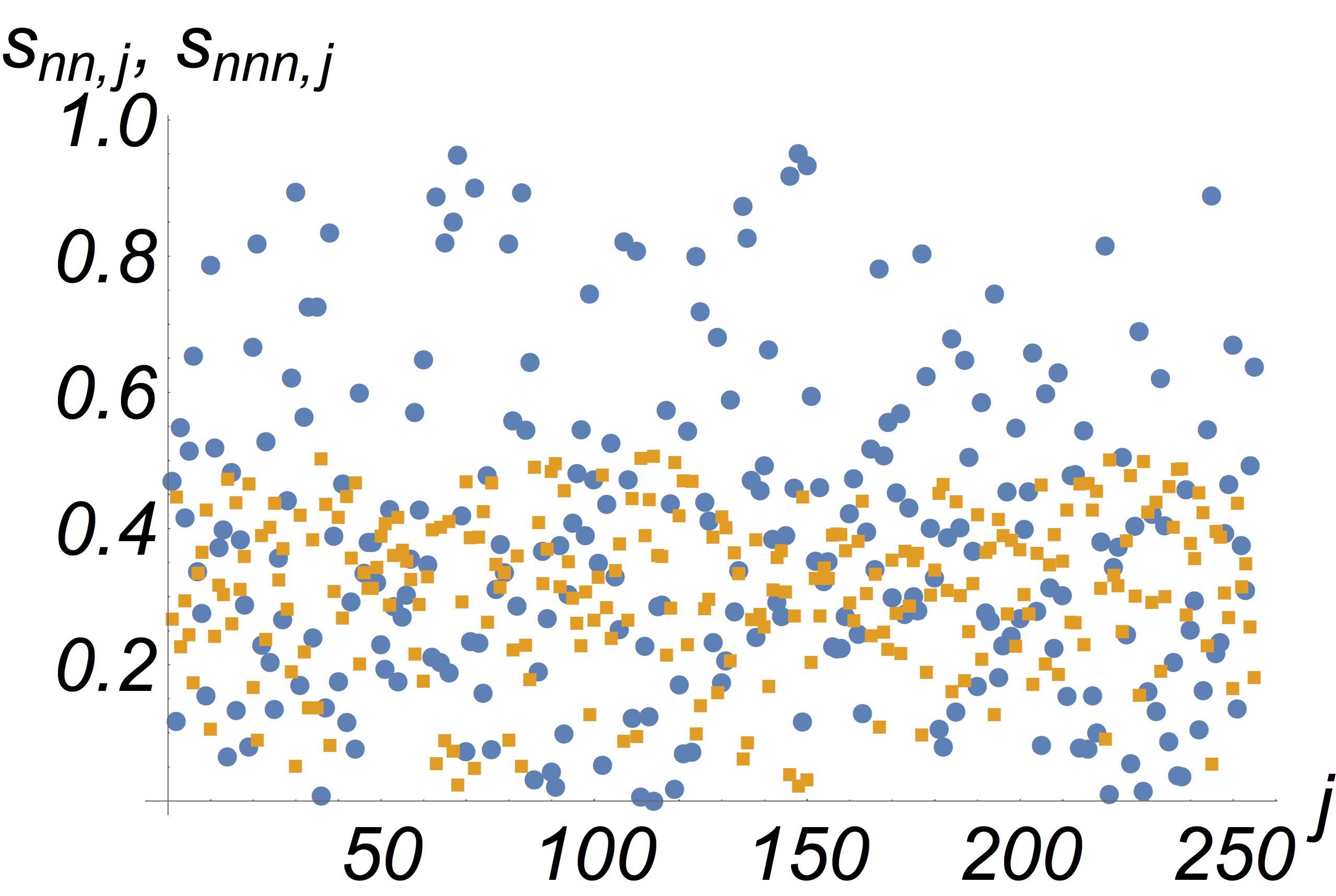}
\\
~~~~
\\
\includegraphics[width=0.40\textwidth]{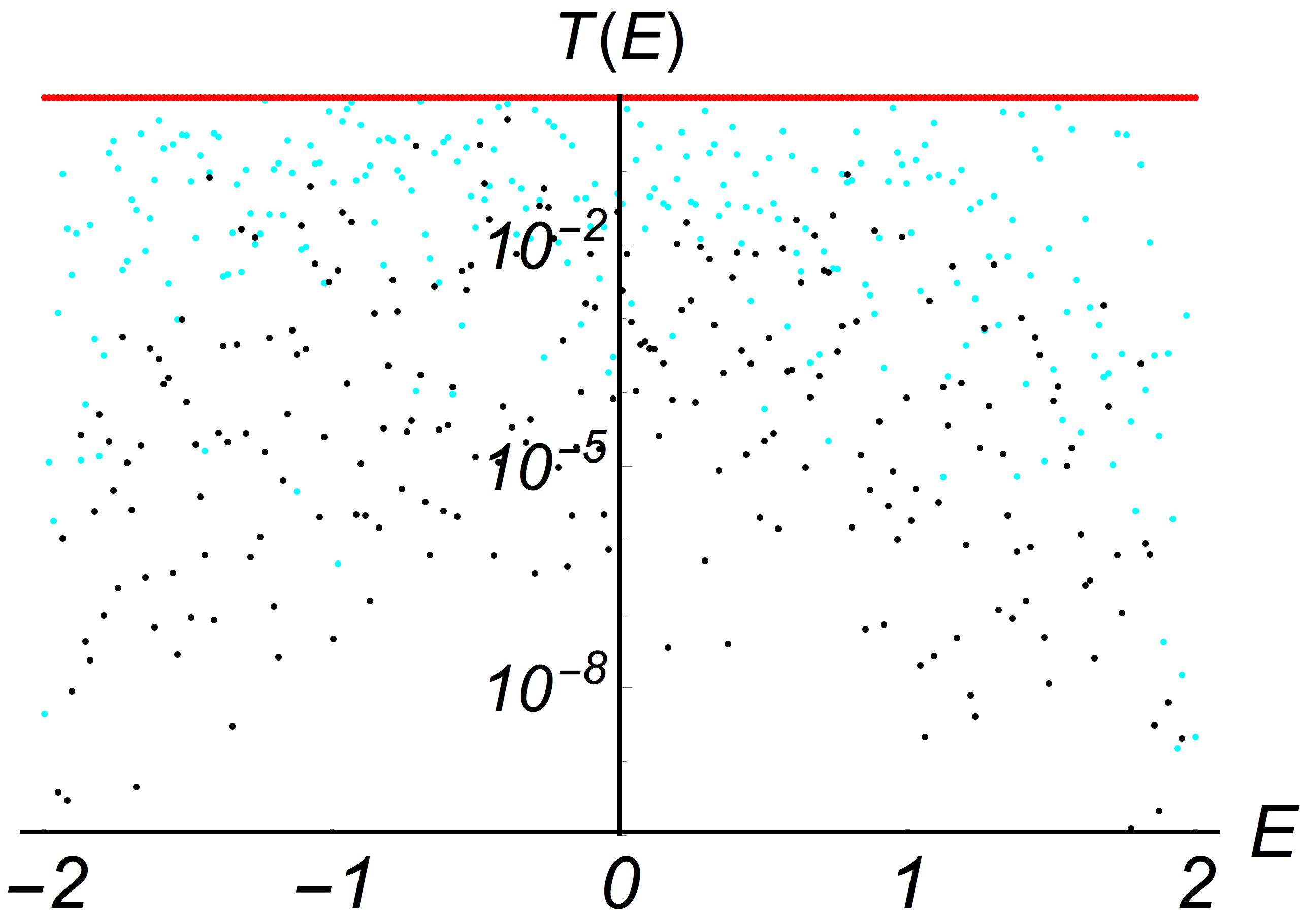}
\caption{
\label{Fig:SQdragon:Disorder}
(Color online.) 
An example of uniform leads connected to a disordered, rectangular nanodevice
with $\ell$$=$$256$ and $m$$=$$16$.  
(Top) The correlated random values for slice $j$ for $\epsilon_j$ (blue circles) 
and $t_j$ (orange squares).  
(Middle) The correlated random values between slices numbered $j$ and $j\!+\!1$ 
for $s_{nn,j}$ (blue circles) and $s_{nnn,j}$ (orange squares).  
(Bottom) Transmission as a function of energy for the disorder in the two upper 
graphs (red), showing the quantum dragon condition ${\cal T}(E)=1$.  The three 
values shown are $\Delta$$=$$0$ (red) for correlated disorder only, and 
for added on~site uncorrelated disorder of strength 
$\Delta=0.25$ (cyan) and $\Delta=0.5$ (blue).  
}
\end{figure}
\end{center}

Here we provide a numerical example to illustrate the concept of quantum dragons for uniform leads,  
based on the analysis in the previous section.  The nanodevice can be viewed as related to the one in 
Fig.~\ref{Fig:SQdragon:Fig02}, except for larger $m$ and $\ell$ values.  
Our numerical results are shown in Fig.~\ref{Fig:SQdragon:Disorder}.  

Any distribution for picking the random TB parameters is allowed.  
For Fig.~\ref{Fig:SQdragon:Disorder} the intra-slice terms, $t_j$, were 
uniformly distributed between $[0,2 t_{lead}]$ and in this section we set $t_{lead}=1$.  
Then we set all $m$ on site energies of the atoms in slice $j$ to be given by 
Eq.~(\ref{Eq:S4:Adragon}).  
The top plot in Fig.~\ref{Fig:SQdragon:Disorder} shows our explicit values 
for $\epsilon_j$ (blue circles) and $t_j$ (orange circles).  
In order to keep all hopping strengths real and positive, since we are studying devices 
in zero magnetic field, for the inter-slice TB parameters for every slice $j$ 
we choose two random numbers $r_{nn,j}$ and $r_{nnn,j}$ uniformly distributed in 
$[0,2t_{lead}]$ and then to satisfy Eq.~(\ref{Eq:S4:Bdragon}) choose the 
inter-slice hopping strengths to be 
\begin{equation}
\label{Eq:dragon:InterTuneNN}
\begin{array}{lcl}
s_{nn,j} & = & 
\frac{r_{nn,j} \> t_{lead}}{r_{nn,j}+2 r_{nnn,j}\cos\left(\frac{\pi}{m+1}\right)} \\
s_{nnn,j} & = & 
\frac{r_{nnn,j} \> t_{lead}}{r_{nn,j}+2 r_{nnn,j}\cos\left(\frac{\pi}{m+1}\right)} \\
\end{array}
\end{equation}
with the explicit values used shown in Fig.~\ref{Fig:SQdragon:Disorder}.  

In Fig.~\ref{Fig:SQdragon:Disorder}, 251 energies equally distributed between 
$-1.999$$\le$$E$$\le$$1.999$ were calculated.  There are singularities present at 
the lead band edges at $E=\pm 2$, so these were avoided.  
For every energy $E$ the matrix ${\bf M}_{256}$, with the same structure 
as the matrix ${\bf M}_4$ in Eq.~(\ref{EQ:S2:Tlarge}), was numerically inverted.  
Since $\ell m=4096$ the matrix ${\bf M}_{256}$ is of dimension $4098$$\times$$4098$.  
As expected for the quantum dragon system, in Fig.~\ref{Fig:SQdragon:Disorder} we 
see ${\cal T}(E)$$=$$1$ for all $E$ (red circles, which run together 
so they seem to form a line segment).  

We also wanted to see how ${\cal T}(E)$ behaves when we move away from the 
condition of correlated disorder.  For uncorrelated disorder, due to Anderson 
localization \cite{ANDE1958,SCHW2007}, we expect very small ${\cal T}(E)$ for almost all 
energies for a finite device size (finite $\ell m$).  
Therefore we also calculated ${\cal T}(E)$ for the case where to the on site 
energy $\epsilon_j$ we added a small uncorrelated random value which is different 
for all $\ell m$ atoms in the nanodevice.  Explicitly, we chose a random variate 
from a normal distribution of mean zero and standard deviation unity, and then multiply by 
a strength $\Delta$. The same random numbers were used for all values of $\Delta$ studied.  
Note if $\Delta$$\ne$$0$ no exact mapping onto an equivalent 1D system is known.  
Figure~\ref{Fig:SQdragon:Disorder} shows results for the three values 
$\Delta$$=$$0$ (a quantum dragon, red points), $\Delta$$=$$0.25$ (cyan points), 
and $\Delta$$=$$0.5$ (blue points).  As expected for $\Delta$$\ne$$0$ the transmission is 
almost always small due to Anderson localization, 
and is typically smaller for larger values of $\Delta$.  
Note the logarithmic scale in Fig.~\ref{Fig:SQdragon:Disorder} for ${\cal T}(E)$, 
so the device has become an insulator for $\Delta$$\ne$$0$, while it is a quantum 
dragon with a perfectly conducting channel for $\Delta$$=$$0$.  

\section{\label{Sec:05}Discussion and conclusions}

We have found the quantum dragon property, namely a perfectly conducting channel, 
${\cal T}(E)$$=$$1$, for nanostructures based on rectangular graphs.  The graphs 
have open boundary conditions, with two sides connected to semi-infinite leads.  
The hopping is nn, and between slices nnn hopping is also included.  For no disorder, 
the graphs become a rectangular crystal, and band structure properties can be 
determined.  When there is strong disorder, as in Fig.~\ref{Fig:SQdragon:Fig02} 
and Fig.~\ref{Fig:SQdragon:Fig04}, band structure is ill defined.  Nevertheless, even though 
there is arbitrarily strong scattering, ${\cal T}(E)$$=$$1$ for all energies which propagate 
through the leads.  Hence quantum dragon nanodevices can be based on 
rectangular graphs.  For the quantum dragon nanodevices, because ${\cal T}(E)$$=$$1$,
the electrical conductance will be $G$$=$$G_0$ in two terminal and $G$$=$$\infty$ 
in four terminal measurements.  Furthermore, the shot power noise is $P$$=$$0$.  

The existence of quantum dragon solutions for electron transmission is extremely relevant 
because of 
recent experimental single layer thick materials based on rectangular lattices which have 
been recently synthesized, including monolayers of Fe \cite{ZHAO2014} 
and of CuO \cite{YIN2016,KANO2017}.

Future work includes finding quantum dragons for different boundary conditions of 
the underlying rectangular lattices.  
Some of these were included in a recent Ph.D.\ dissertation \cite{INKO2017}.  
In order to find quantum dragons with strong disorder, the crucial idea is all 
intra-slice and all inter-slice parts of the Hamiltonian must have a common 
eigenvector, and furthermore this eigenvector must correspond to a physically 
realizable connection to the leads.  For nanodevices based on rectangular lattices, 
this is possible for other boundary conditions \cite{INKO2017}.
These studies may benefit from renormalization group calculations for transport, 
such as have been used for hierarchical lattices \cite{Hanoi2011}.  

\begin{center}
\begin{figure}[tb]
\vspace{0.05 cm}
\includegraphics[width=0.45\textwidth]{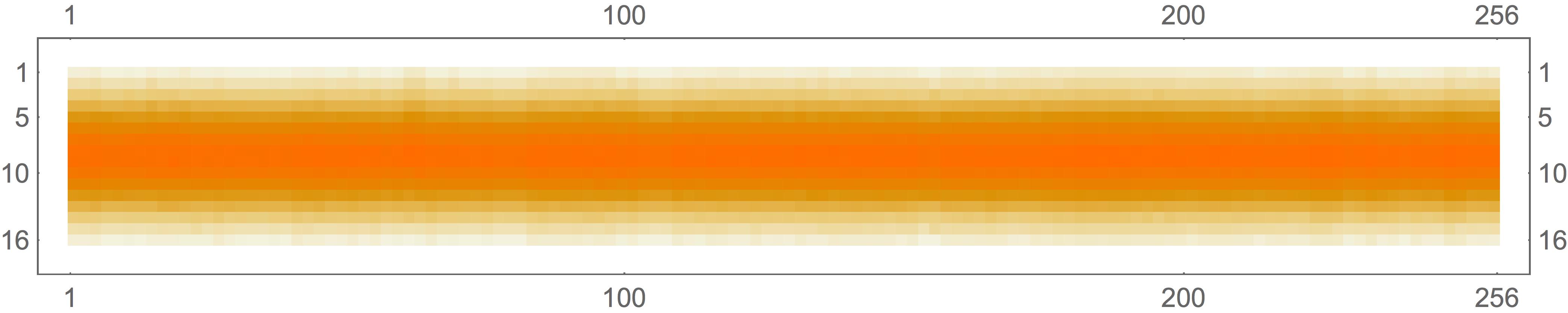}
\\
~~~~
\\
\includegraphics[width=0.45\textwidth]{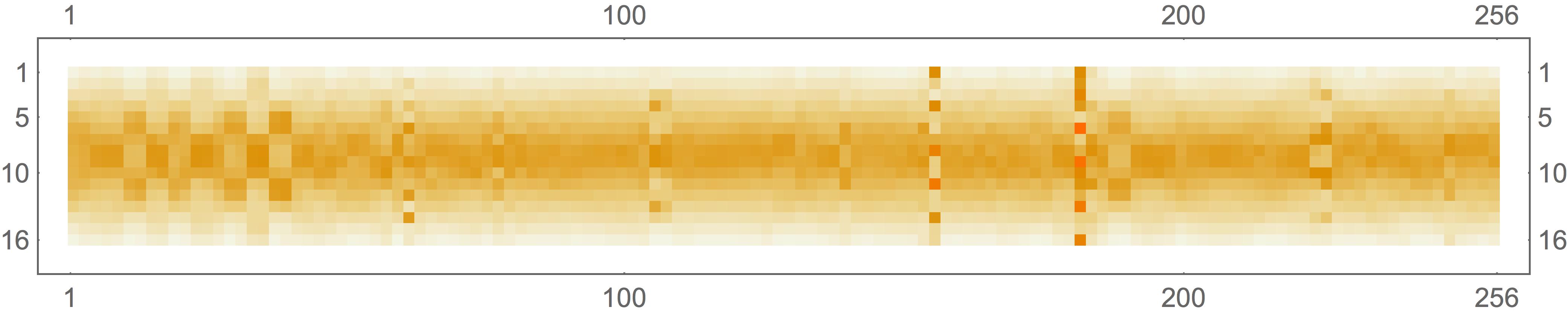}
\\
~~~~
\\
\includegraphics[width=0.45\textwidth]{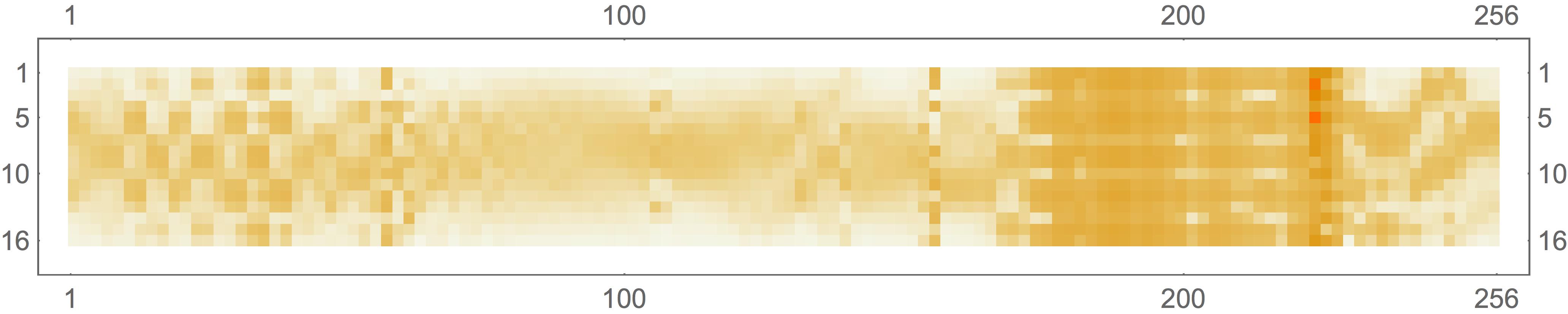}
\caption{
\label{Fig:LDOS}
(Color online.) 
The local density of states (LDOS) for a strongly disordered 
nanodevice based on a rectangular lattice with  
$\ell$$=$$256$ and $m$$=$$16$, attached to uniform leads.  
The correlated disorder in the TB parameters 
$\epsilon_j$, $t_j$, $s_{nn,j}$, and $s_{nnn,j}$ is similar to 
Fig.~\ref{Fig:SQdragon:Disorder}.  
These are calculated with $E$$=$$0.01 t_{lead}$, but the results are similar for other energies.  
The color coding has a larger LDOS for brighter, more orange pixels.
(Top) 
The LDOS is independent of the slice $j$ when there is only correlated disorder, 
so the device is a quantum dragon.  Even though there is strong scattering, ${\cal T}$$=$$1$.  
(Middle) With added uncorrelated on site disorder of strength $\Delta$$=$$0.01$ the LDOS changes
dramatically, while the transmission decreases to ${\cal T}$$=$$0.99851$.
(Bottom) The LDOS changes even more when for the same uncorrelated on site random disorder the 
strength is $\Delta$$=$$0.02$, while the transmission has further decreased to 
${\cal T}$$=$$0.98205$. 
}
\end{figure}
\end{center}

Other further work could be to study the even-odd structure of rectangular CuO lattices, 
where experimentally even-numbered and odd-numbered slices have different 
structures \cite{YIN2016,KANO2017}. One expects again quantum dragon solutions, because the 
same type of even-odd structure was exploited in the study of quantum dragon solutions 
in single walled carbon nanotubes \cite{MANdragon2014}.  
Further work can also be performed examining quantities such as a local density of states (LDOS).  
The Green's function ${\cal G}(E)$ is given in Eq.~(\ref{Eq:AppD:09}), 
and the LDOS at $E$ is the imaginary part of the diagonal elements of ${\cal G}(E)$ 
divided by $\pi$ \cite{SOUM2002}.  
For a quantum dragon, as shown in 
Fig.~\ref{Fig:LDOS}, the correlated disorder of a quantum dragon gives a LDOS 
independent of the slice index $j$, even though the disorder on every slice 
is different.  

A further investigation of added disorder near a quantum dragon solution is also 
warranted.  We have observed such an analysis will require the study of the Fano 
resonances \cite{MIRO2010} present, which are a source of the small values for electron 
transmission seen in Fig.~\ref{Fig:SQdragon:Disorder} and \ref{Fig:SQdragon:Disorder:Dimerized}.  
The full counting statistics of nanodevices \cite{FLIN2008} nearby in parameter space 
to quantum dragon solutions would also be of interest.  
One could also investigate how such quantum dragon solutions behave if one goes beyond 
the TB model, for example using an exact discretization of the Schr{\"o}dinger 
equation \cite{TARA2016}, or due to many-electron effects such as were used recently 
to analyze transport through a nanoscale ring-dot device \cite{BIBO2016}.  

The possibility of technologically using quantum dragon solutions for field effect transistors 
or sensors \cite{MANpatent} also deserves further explaination.  Furthermore, the possibilities 
exist related dragon solutions may also be present in other strongly disordered systems, 
for example where the open boundary conditions and disorder would normally lead to transverse 
Anderson localization \cite{SCHW2007} in optical waveguides.

\section*{Acknowledgements}
One of the authors (MAN) thank Tom{\'a}{\v s} Novotn{\' y} and Maciej Ma{\'s}ka for useful discussions, 
and the Faculty of Mathematics and Physics at 
Charles University in Prague, Czech Republic for hospitality during 
a stay as a Fulbright Distinguished Chair.  Funding for MAN as a Fulbright Distinguished Chair 
is gratefully acknowledged.


\appendix
\section*{\label{AppA} Appendix A: FOR FIGURES WITHOUT DISORDER}

The complete description of two figures of the device and leads 
without disorder, namely Fig.~\ref{Fig:SQdragon:Fig01} and 
Fig.~\ref{Fig:SQdragon:Fig03} are given.  

\subsection{\label{AppAsub1} Figure~\ref{Fig:SQdragon:Fig01} complete description}

An example of the case of uniform (non-dimerized) leads, and no disorder in a rectangular lattice 
device is shown in Fig.~\ref{Fig:SQdragon:Fig01}.  
Because of the lack of disorder, the device can be considered to be a planar, rectangular crystal.  
The location of the device is highlighted in light yellow.  
The vertical gray lines show the division into slices, both for the device and 
for the lead atoms.  
Here there are $\ell$$=$$16$ slices in the device, and $m$$=$$7$ atoms in each slice.   
Therefore, the device has $m\ell$$=$$112$ atoms (red spheres).  
The intra-slice hopping terms are shown by the vertical line segments 
(green), TB parameters $t_j$$=$$t$ for slice $j$, and are only between nn atoms.  
The inter-slice hopping terms are for nn interactions shown by the 
horizontal line segments (red-orange, TB parameter $s_{nn,j}$$=$$s_{nn}$), 
and the nnn interactions shown by the 
(black, TB parameter $s_{nnn,j}$$=$$s_{nnn}$) line segments which form an X-shape.  
Only four atoms (blue-green) for both the incoming and outgoing semi-infinite leads are shown.  
The connections between the leads and the device are shown by line segments (cyan) 
with the width proportional to the elements of the eigenvector ${\vec x}_1$ in Eq.~(\ref{EQ:S3:Aeigenvectors}).

\subsection{\label{AppAsub2} Figure~\ref{Fig:SQdragon:Fig03} complete description}

Fig.~\ref{Fig:SQdragon:Fig03} shows 
an example of the case of dimerized leads and a dimerized device, 
for no disorder in the rectangular device.
Consequently, the device may be considered to be planar.  
The device is highlighted in light yellow.  
The vertical gray lines show the division into slices, both for the device and 
for the lead atoms.  
Here there are $\ell$$=$$4$ slices in the device, and $m$$=$$8$ atoms in each slice.   
Therefore, the device has $m\ell$$=$$32$ atoms, denoted by different sized (and colored) 
spheres for the odd slices (red) and even slices (green). 
The different colors and sizes represent different values for the TB on site energies $\epsilon_j$.  
The first (leftmost) slice is numbered one.  
The intra-slice hopping terms are only between nn atoms, and 
are shown by the vertical line segments which are thick and green for the 
odd-numbered slices and thinner and yellow for the even-numbered slices, representing TB parameters $t_j$.  
The inter-slice hopping terms are for nn interactions shown by the 
horizontal line segments (magenta for odd-to-even and 
yellow for even-to odd, representing TB parameters $s_{nn,j}$), and the nnn interactions shown by the 
line segments which form an X-shape (magenta for odd-to-even and 
black for even-to-odd, representing the TB parameters $s_{nnn,j}$).  
Only five atoms for the incoming lead and four atoms for the outgoing semi-infinite lead are shown
(large and fuchsia colored for odd-numbered atoms, while small and black for 
even-numbered atoms, representing $\epsilon_o$ and $\epsilon_e$ TB parameters, respectively).  
The lead hopping interactions are shown as thick (white) cylinders for 
odd-to-even hopping terms (representing TB parameters $t_{oe}$), and 
thinner (yellow) cylinders for even-to-odd hopping terms (representing TB parameters $t_{eo}$).  
The connections between the leads and the device are shown by line segments 
(cyan for the incoming lead, and magenta for the outgoing lead) 
with the cylinder width proportional to the elements of the eigenvector ${\vec x}_1$ in 
Eq.~(\ref{EQ:S3:Aeigenvectors}).

\appendix
\section*{\label{AppB} Appendix B: FIGURES  WITH DISORDER}
The description for Fig.~\ref{Fig:SQdragon:Fig02} and Fig.~\ref{Fig:SQdragon:Fig04} is presented.  

Consider a model where atoms in slice $j  $ must be the same distance $a_j$ apart, as in a 
ball-and-stick polymer model.  
Hence, between nn atoms within slice $j$ there is only a single 
distance $a_j$ allowed between the atoms (equivalent to attachment of springs with 
infinite spring constant and equilibrium 
distance $a_j$).  

Between atoms in two neighboring slices, there are nn and nnn 
equilibrium spring distances.  
Let the hard spheres in slice $j$ be of radius $R_j$.  
Follow Konishi, et al \cite{Koni2008} for the Monte Carlo simulations and 
energy functions.
The nn elastic interactions are
\begin{equation}
\label{Eq:AppB:01}
{\cal H}_{nn} = \frac{m\> k_{j,j+1}}{2} 
\left(r_{j,j+1}-\left(R_j+R_{j+1}\right)\right)^2 
\end{equation}
since all distances must be the same between the 
intra-slice nn atoms.  Here 
\begin{equation}
\label{Eq:AppB:02}
\begin{array}{lcl}
r_{j,j+1} & = &
\sqrt{\left(x_{j,i}-x_{j+1,i}\right)^2+\left(y_{j,i}-y_{j+1,i}\right)^2} \\
& = & \sqrt{\Delta x_{j}^2+\Delta y_{j}^2}
\end{array}
\end{equation}
is the distance (with the atoms confined to have the same $z$ value (with the $z$ 
value the direction along which current will flow).  
Here the index $i$ labels the $m$ atoms in the slice labeled $j$.  
This distance must be the same 
between nn for all atoms in slices $j$ and $j+1$, and consequently, they must have the same value of 
$\Delta x_{j}$ and $\Delta y_{j}$.  
Similarly, the nnn elastic term is 
\begin{equation}
\label{Eq:AppB:03}
{\cal H}_{nnn} = \frac{2 (m-1) \> {\hat k}_{j,j+1}}{2} 
\left({\hat r}_{j,j+1}-\sqrt{2}\left(R_j+R_{j+1}\right)\right)^2 
\end{equation}
with
\begin{equation}
\label{Eq:AppB:04}
\begin{array}{lcl}
{\hat r}_{j,j+1} & = &
\sqrt{\left(x_{j,i}-x_{j+1,i\pm 1}\right)^2+\left(y_{j,i}-y_{j+1,i\pm 1}\right)^2} \\
& = & \sqrt{\Delta {\hat x}_{j}^2+\Delta {\hat y}_{j}^2}
\>.
\end{array}
\end{equation}

We assume the hard sphere radii (the radii of the plotted spheres in the figures) is proportional to 
$|\epsilon_j|$ for slice $j$, namely $R_j\propto |\epsilon_j|$.  
We assume the distance between atoms within a slice has $a_j\propto t_j$, with $t_j$ also reflected 
by the radii of the cylinders representing the intra-slice bonds in the figures.  
We assume the spring constants between slices have 
$k_{j,j+1}\propto s_{nn,j}$ and ${\hat k}_{j,j+1}\propto s_{nnn,j}$, and also the 
width of the cylinders 
representing the bonds have the same proportionality.  One then has a complete (classical) 
Hamiltonian for the nanosystems we study.  The configuration of the classical representation of the 
nanodevice is then found by perfoming a simulated annealing Monte Carlo process to attempt to 
minimize the total elastic energy of the nanodevice.  

\subsection{\label{AppAsub3} Figure~\ref{Fig:SQdragon:Fig02} complete description}
Figure~\ref{Fig:SQdragon:Fig02} contains an example of a nanodevice with uniform 
(non-dimerized) leads.  The device has $\ell$$=$$20$ slices, and $m$$=$$8$ atoms in every slice.  
The underlying graph is rectangular, with nn interactions and also with nnn interactions 
between atoms in neighboring slices.  
Only five atoms in both the incoming and outgoing semi-infinite leads are shown.  
The radii of the spheres are proportional to $|\epsilon_j|$, and the radii of the 
cylinders representing the bonds are proportional to the hopping strengths.  
The bonds are magenta for nnn bonds (TB parameter $s_{nnn,j}$ for slice $j$), 
cyan for nn inter-slice bonds (TB $s_{nn,j}$), and green for nn 
intra-slice bonds (TB $t_j$). 
The $t_j$ are different for every slice, 
and were chosen to be uniformly distributed in $[0,2 t_{lead}]$.  
The on site energies of slice $j$ are all the same, with the $\epsilon_j$ given by the 
quantum dragon condition of Eq.~(\ref{Eq:S4:Adragon}).  
The inter-slice bonds are also all different, with the $s_{nn,j}$ and 
$s_{nnn,j}$ chosen to satisfy the quantum dragon condition of Eq.~(\ref{Eq:S4:Bdragon}).  
The connections between the leads and the device, blue-green cylinders, have strengths  
given by Eq.~(\ref{Eq:S4:wudragon}).  
Note the extreme disorder in the system, and that it is very far from a rectangular crystal, 
rather it is a structure based on a rectangular graph.  Nevertheless, for transmission it 
has ${\cal T}(E)$$=$$1$.  

\subsection{\label{AppAsub4} Figure~\ref{Fig:SQdragon:Fig04} complete description}
Figure~\ref{Fig:SQdragon:Fig04} contains an example of a nanodevice with dimerized leads.  
The device has $\ell$$=$$16$ slices, and $m$$=$$7$ atoms in every slice.  
The underlying graph is rectangular, with nn interactions and also with nnn interactions.  
Only five atoms in both the incoming and outgoing semi-infinite leads are shown, 
with different colors and radii reflecting different on site energies, 
$\epsilon_o$ (pink) and $\epsilon_e$ (black).  The hopping terms in the leads 
are also dimerized, representing $t_{eo}$ (cyan, thick cylinder, smaller lattice spacing) and 
$t_{oe}$ (pink, thin cylinder, larger lattice spacing).  
The radii of the spheres in the nanodevice are proportional to $|\epsilon_j|$, 
with atoms red for odd-numbered $j$ and black for even-numbered $j$.  
The radii of the 
cylinders representing the bonds are proportional to the hopping strengths.  
The nnn bonds form an X-shape (TB parameter $s_{nnn,j}$ for slice $j$, magenta for odd $j$, 
cyan for even $j$).  
The nn inter-slice bonds, TB $s_{nn,j}$, are also magenta for odd $j$ and 
cyan for even $j$.  
The nn intra-slice bonds, TB $t_j$, are green for $j$ odd and yellow for $j$ even.  
The $t_j$ are different for every slice, 
and were chosen to be uniformly distributed in $[0,2 t_{oe}]$ ($[0,2 t_{eo}]$) for 
odd (even) $j$.  
The on site energies of slice $j$ are all the same, with the $\epsilon_j$ given by the 
quantum dragon condition of Eq.~(\ref{Eq:AD:S3:kappaDragon2}).  
The inter-slice bonds are also all different, with the $s_{nn,j}$ and 
$s_{nnn,j}$ chosen to satisfy the quantum dragon condition of Eq.~(\ref{Eq:Norm2DnnOnnn}) 
and Eq.~(\ref{Eq:Norm2DnnEnnn}).  
The connections between the leads and the device, blue-green cylinders, have strengths  
proportional to ${\vec x}_1$ with strengths given by the dragon condition 
in Eq.~(\ref{EQ:AD:S3:wANDuDRAGON}).  
Note the extreme disorder in the system, and that it is very far from a rectangular crystal, 
being rather based on a rectangular graph.  Nevertheless, for transmission it 
is a quantum dragon with ${\cal T}(E)$$=$$1$ for all energies that propagate in the leads.  

\appendix
\section*{\label{AppC} Appendix C: Derivation of matrix method for dimerized leads}
In this appendix, an outline of the derivation for the transmission via the matrix 
method is given.  
This parallels Appendix~A of ref.~\onlinecite{MANdragon2014}, with the exception here different 
on~site energies for the dimerized leads are also included.  
This parallels the case of uniform leads \cite{DCA2000}, which is also 
derived in the supplemental material of the 2014 quantum dragon paper \cite{MANdragon2014}.  
For uniform leads the matrix form in the main text is obtained from results in this 
section.  

\begin{center}
\begin{figure}[tb]
\vspace{0.05 cm}
\includegraphics[width=0.30\textwidth]{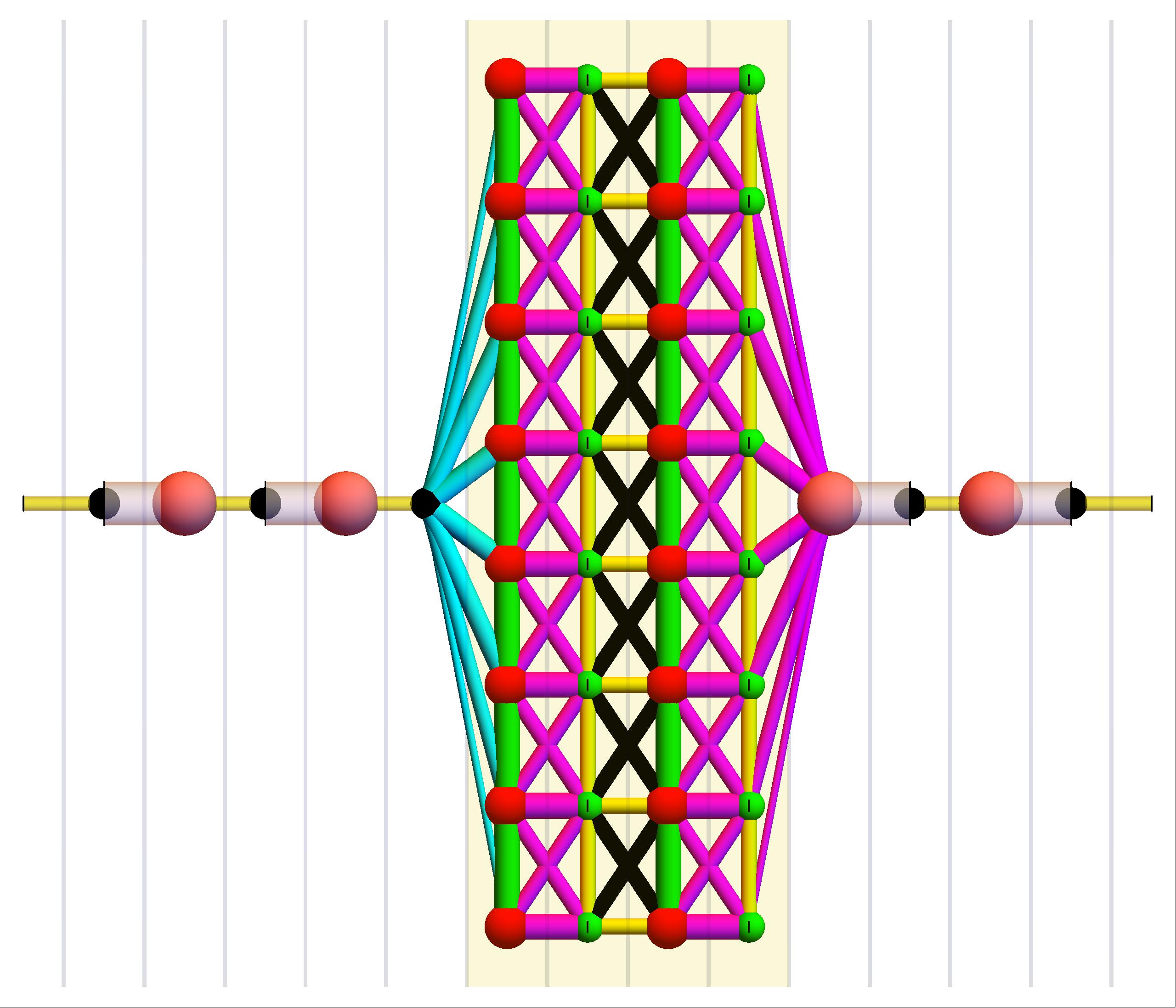},
\caption{
\label{Fig:SQdragon:Fig03}
(Color online.) 
An example of dimerized leads connected to a 
dimerized rectangular device, both without disorder.  
See Appendix~A.2 
for a complete description.  
}
\end{figure}
\end{center}

Consider a nanodevice with $m$ atoms in every slice and with $\ell$ slices, connected 
to dimerized leads, with an example shown in Fig.~\ref{Fig:SQdragon:Fig03}.  
The leads (both incoming and outgoing) have on~site energy 
$\epsilon_e$ ($\epsilon_o$) for even (odd) numbered sites.  The arrangement, including numbering 
of the lead atoms, is as in the figure in Appendix~A of ref.~\onlinecite{MANdragon2014}.  
The outgoing lead is first completely analyzed, then the incoming lead is addressed.  The 
traveling-wave ansatz within the outgoing lead is the same as 
Eq.~[A3] of ref.~\onlinecite{MANdragon2014} (note equation numbers 
from the 2014 paper \onlinecite{MANdragon2014} are enclosed in square brackets)
\begin{equation}
\label{Eq:AppC:01}
\begin{array}{lclcl}
\psi_j & \> = \> & t_T \>\> e^{i q(j-1)} & \qquad\qquad & j=1,\>3,\> 5,\cdots
\\
\psi_j & \> = \> & t_T \delta e^{i q(j-1)} & \qquad\qquad & j=2,\> 4,\> 6,\cdots
\>.
\end{array}
\end{equation}
\begin{widetext}
The ansatz makes use of Bloch's theorem
\cite{Bloch} through the phase factor $\delta$.  Multiplying through 
by the Hamiltonian of the outgoing lead becomes 
(corresponding to Eq.~[A4] of ref.~\onlinecite{MANdragon2014})
\begin{equation}
\label{Eq:AppC:02}
\begin{array}{lclcl}
t_T e^{i q (j-1)} \left[-t_{eo} \delta e^{-i q}+\left(\epsilon_o-E\right)-t_{oe}\delta e^{iq}\right] 
& \> = \> 0 & \qquad &  j=3,\> 5,\> 7,\cdots
\\
t_T e^{i q (j-1)} \left[-t_{oe} e^{-i q}+\left(\epsilon_e-E\right)\delta-t_{eo} e^{iq}\right] 
& \> = \> 0 & \qquad &  j=2,\> 4,\> 6,\cdots
\end{array}
\end{equation}
\end{widetext}
which can be solved to eliminate $\delta$ to give (corresponding to Eq.~[A5]) 
\begin{equation}
\label{Eq:AppC:03}
\cos(2 q) \> = \> \frac{\left(\epsilon_e-E\right)\left(\epsilon_o-E\right)-t_{eo}^2-t_{oe}^2}{2 t_{eo}t_{oe}}
\end{equation}
and using the double angle formula for $\cos(2q)$ gives
\begin{equation}
\label{Eq:AppC:04}
\cos(q) \> = \> \pm \sqrt{\frac{\left(\epsilon_e-E\right)\left(\epsilon_o-E\right)
-\left(t_{eo}-t_{oe}\right)^2}{4 t_{eo}t_{oe}}}
\end{equation}
corresponding to Eq.~[A6].  
A propagating wave requires $q$ to be real, and hence $0\le\cos(q)\le 1$ or
$-1\le\cos(q)\le 0$ for the two signs in front of the square root in Eq.~(\ref{Eq:AppC:04}).  

One is free to set the zero for energy for the entire system, and a reasonable choice 
is to set the zero at $\left(\epsilon_o+\epsilon_e\right)/2$.  
This can be accomplished by insisting 
$\epsilon_o=-\epsilon_e$, with the zero of energy then at zero, 
which is also 
the midpoint between $\epsilon_o$ and $\epsilon_e$.  
This gives 
propagating waves (corresponding to Eq.~[A7]) for positive energies of
\begin{widetext}
\begin{equation}
\label{Eq:AppC:05}
\sqrt{\left(\frac{\epsilon_e-\epsilon_o}{2}\right)^2+\left(t_{eo}-t_{oe}\right)^2}
\le E\le 
\sqrt{\left(\frac{\epsilon_e-\epsilon_o}{2}\right)^2+\left(t_{eo}+t_{oe}\right)^2}
\end{equation}
and
\begin{equation}
\label{Eq:AppC:06}
-\sqrt{\left(\frac{\epsilon_e-\epsilon_o}{2}\right)^2+\left(t_{eo}+t_{oe}\right)^2}
\le E\le 
-\sqrt{\left(\frac{\epsilon_e-\epsilon_o}{2}\right)^2+\left(t_{eo}-t_{oe}\right)^2}
\end{equation}
\end{widetext}
for negative energies.  
By manipulating the two expressions in Eq.~(\ref{Eq:AppC:02}) one 
finds 
\begin{equation}
\label{Eq:AppC:07}
\begin{array}{lcl}
\delta & = & \frac{\epsilon_o-E}{t_{oe}e^{i q}+t_{eo}e^{-iq}}
\\
& = & \frac{t_{eo}e^{iq}+t_{oe}e^{-i q}}{\epsilon_e-E}
\end{array}
\end{equation}
corresponding to Eq.~[A8].  
One also finds using the outgoing lead equations
as in Eq.~[A10]
\begin{equation}
\label{Eq:AppC:08}
\begin{array}{lcl}
\xi_u & = & \epsilon_o-E-t_{oe}\delta e^{iq}
\\
& = & \left(\epsilon_o-E\right)\left(1-t_{oe}\frac{e^{i q}}{t_{oe}e^{i q}+t_{eo}e^{- i q}}\right)
\\
& = & \left(\epsilon_o-E\right) \frac{t_{eo} e^{- i q}}{t_{oe} e^{i q}+t_{eo}e^{- i q}}
\end{array}
\end{equation}
corresponding to Eq.~[A12].  

For the incoming lead one obtains
via the traveling-wave ansatz 
\begin{widetext}
\begin{equation}
\begin{array}{lclclcl}
e^{-i q j}\left[\epsilon_o-E-t_{eo}\delta e^{- i q}-t_{oe}\delta e^{i q}\right]
& + & 
r e^{i q j} \left[\epsilon_o-E-t_{eo}\delta^*e^{i q}-t_{oe}\delta^*e^{- i q}\right]
& \> = \> & 0 
& \qquad\qquad & j=1,3,5,\cdots 
\\
e^{- i q j} \left[\delta\left(\epsilon_e-E\right)-t_{oe}e^{- i q}-t_{eo} e^{i q}\right]
& + &
r e^{i q j} \left[\delta^*\left(\epsilon_e-E\right)-t_{oe}e^{i q}-t_{eo}e^{- i q}\right]
& = & 0
& & j=2,4,6 \cdots
\end{array}
\end{equation}
\end{widetext}
which has all four terms in square brackets zero for 
the expressions for $\delta$ in Eq.~(\ref{Eq:AppC:07}).  
When $\epsilon_o\ne\epsilon_e$, the values of $\xi$ are different for the 
incoming ($\xi_w$) and outgoing ($\xi_u$) leads.  
The incoming lead requires the association 
\begin{equation}
\label{Eq:AppC:09}
\begin{array}{lcl}
\xi_w & = & 
\epsilon_e-E-\frac{1}{\delta^*}t_{oe} e^{i q} \\
& = &  \left(\epsilon_e-E\right) \frac{t_{eo} e^{- i q}}{t_{oe}e^{i q} + t_{eo} e^{- i q}}
\end{array}
\end{equation}
with the further definition
\begin{equation}
\label{Eq:AppC:10}
\begin{array}{lcl}
\Lambda & = & 
\xi_w\delta -\left(\epsilon_e-E\right)\delta+t_{oe}e^{- i q}
\\
& = & 
-t_{oe}\left(\frac{\delta}{\delta^*}e^{i q}-e^{- i q}\right)
\>.
\end{array}
\end{equation}
Consequently, one obtains the solution to the transmission ${\cal T}=\left|t_T\right|^2$ from the 
solution to the matrix equation of the form (written for $\ell$$=$$4$) 
\begin{widetext}
\begin{equation}
\label{EQ:AD:S2:Tlarge}
{\bf M}_4 
\left(\begin{array}{c}
\! \delta + r \delta^* \! \\
{\vec \psi}_1 \\
{\vec \psi}_2 \\
{\vec \psi}_3 \\
{\vec \psi}_4 \\
t_T \\
\end{array}\right)
\> = \> 
\left(\begin{array}{cccccc}
\! \xi_w & {\vec w}^\dagger & {\vec 0}^\dagger & {\vec 0}^\dagger & {\vec 0}^\dagger & 0 \\
{\vec w} & {\bf F}_1 & {\bf B}_{12} & {\bf 0} & {\bf 0} & {\vec 0} \\
{\vec 0} & {\bf B}_{12}^\dagger & {\bf F}_2 &  {\bf B}_{23} & {\bf 0} & {\vec 0} \\
{\vec 0} & {\bf 0} & {\bf B}_{23}^\dagger & {\bf F}_3 &  {\bf B}_{34} & {\vec 0} \\
{\vec 0} & {\bf 0} & {\bf 0} & {\bf B}_{34}^\dagger & {\bf F}_4 & {\vec u} \\
0 & {\vec 0}^\dagger & {\vec 0}^\dagger & {\vec 0}^\dagger & {\vec u}^\dagger & \! \xi_u \! \\
\end{array}\right)
\!\!
\left(\begin{array}{c}
\! \delta + r \delta^* \! \\
{\vec \psi}_1 \\
{\vec \psi}_2 \\
{\vec \psi}_3 \\
{\vec \psi}_4 \\
t_T \\
\end{array}\right)
= 
\left(\begin{array}{c}
\!\! \Lambda \!\! \\
\!\! {\vec 0} \!\! \\
\!\! {\vec 0} \!\! \\
\!\!{\vec 0} \!\! \\
\!\! {\vec 0} \!\! \\
\!\! 0 \!\! \\
\end{array}\right)
\end{equation}
\end{widetext}
corresponding to Eq.~[A11].  
If $\epsilon_e$$=$$\epsilon_o$ one has $\xi_w$$=$$\xi_u$$=$$\xi$, 
and therefore obtain the specialized case of Eq.~(\ref{EQ:S2:Tlarge}).  
The transmission is calculated by inverting the matrix ${\bf M}_\ell$ for 
a specific energy, and thereby obtaining ${\cal T}=\left|t_T\right|^2$.  

For dimerized leads with $\epsilon_o$$=$$\epsilon_e$$=$$0$ and $t_{eo}$$\ne$$t_{oe}$, 
$\delta$ has the form from Eq.~[A8] and 
consequently the energy range of propagating electrons is expressed as Eq.~[A7].  
For the case of uniform leads, with $\epsilon_o$$=$$\epsilon_e$$=$$0$ and 
$t_{eo}=t_{oe}=t_{lead}$ one has $\delta$$=$$1$, 
\begin{equation}
\label{Eq:AD:2uniform}
\xi_w=\xi_u=\xi=e^{- i q} = \frac{-E-i\sqrt{4 t_{lead}^2-E^2}}{2 t_{lead}}
\end{equation}
and $\Lambda=- 2 i \sin(q)$ with 
electrons of energy $-2 t_{lead}\le E\le 2 t_{lead}$ propagating  
as in refs.~\onlinecite{DCA2000,MANdragon2014,BOET2011,NOVO2014}.  


\appendix
\section*{\label{AppD} Appendix D: Quantum dragons with dimerized leads}

\begin{center}
\begin{figure*}[tb]
\vspace{0.05 cm}
\includegraphics[width=0.90\textwidth]{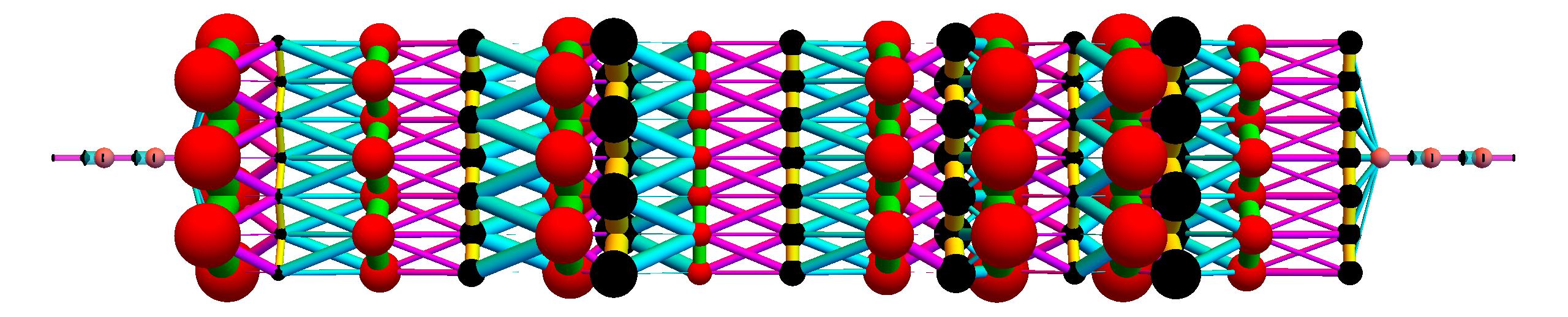}
\\
~~~~
\\
\includegraphics[width=0.90\textwidth]{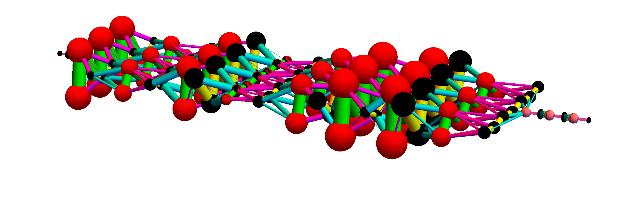}
\caption{
\label{Fig:SQdragon:Fig04}
(Color online.) 
An example of dimerized leads connected to a dimerized, disordered, rectangular device.  
The same device is shown in the top (top view) and 
bottom (oblique view) of the figure.  
See Appendix~B 
for a complete description.  
}
\end{figure*}
\end{center}

In this appendix, the exact mapping as well as the locating of quantum dragons is 
derived for the case of dimerized leads.  We assume an underlying rectangular graph 
with atoms placed on the nodes, and the graph is composed of $\ell$ slices each of 
which have $m$ atoms.  
An example for a rectangular crystal is shown in Fig.~\ref{Fig:SQdragon:Fig03}, 
and an example based on an underlying rectangular graph, but with strong correlated 
disorder is shown in Fig.~\ref{Fig:SQdragon:Fig04}.  

The intra-slice parts of the device Hamiltonian, ${\bf A}_j$ are 
defined in Eq.~(\ref{EQ:S3:DefineA}).  
The inter-slice parts of the device Hamiltonian, ${\bf B}_{j,j+1}$, are 
defined in Eq.~(\ref{EQ:S3:DefineB}).  
For our nanosystem based on an underlying rectangular graph, all ${\bf A}_j$ and 
${\bf B}_{j,j+1}$ are a sum of a constant times the $m$$\times$$m$ identity matrix 
plus another constant times 
the matrix ${\bf Q}$ defined in Eq.~(\ref{EQ:S3:DefineQ}).  
The mutual eigenvector of all ${\bf A}_j$ and ${\bf B}_{j,j+1}$ is the vector 
${\vec x}_1$ defined in Eq.~(\ref{EQ:S3:Aeigenvectors}).  

Define a $m$$\times$$m$ transformation matrix 
\begin{equation}
\label{EQ:AD:S3:DefineX}
\begin{array}{lcl}
{\bf X} & = & \left(\begin{array}{cc} {\vec x}_1 & {\bf Y}_{\rm other}\\ \end{array}\right) \\
& {\rm and} & \\
{\bf X}^\dagger & = & \left(\begin{array}{c} {\vec x}_1^\dagger \\ {\bf Y}_{\rm other}^\dagger \\ 
\end{array}\right) \\
\end{array}
\end{equation}
where ${\bf Y}_{\rm other}$ is a $m$$\times$$(m$$-$$1)$ matrix, which can be thought of as being 
composed of $m$$-$$1$ normalized vectors which are orthogonal to ${\vec x}_1$ and 
also are orthogonal to each other.  
Therefore 
${\bf Y}_{\rm other}^\dagger {\vec x}_1 = {\vec 0}_{m-1}$
and
${\bf Y}^\dagger {\bf Y}={\bf I}_{m-1}$.
Furthermore, 
\begin{equation}
\label{EQ:AD:S3:XYI}
\begin{array}{lcl}
{\bf X}^\dagger {\bf X} 
& \> = \> & 
\left(\begin{array}{c} {\vec x}_1^\dagger \\ {\bf Y}_{\rm other}^\dagger \\ 
\end{array}\right)
\left(\begin{array}{cc} {\vec x}_1 & {\bf Y}_{\rm other}\\ \end{array}\right) 
\\
& \> = \> &  
\left(\begin{array}{cc}
{\vec x}_1^\dagger\cdot{\vec x}_1 & {\vec x}_1^\dagger {\bf Y}_{\rm other} \\
{\bf Y}_{\rm other}^\dagger {\vec x}_1 & {\bf Y}_{\rm other}^\dagger {\bf Y}_{\rm other} \\
\end{array}\right) \\
& \> = \> & 
\left(\begin{array}{cc}1 & {\vec 0}_{m-1}^\dagger \\ {\vec 0}_{m-1} & {\bf I}_{m-1} \end{array}\right) \\
\end{array}
\end{equation}
so 
${\bf X}^\dagger{\bf X} = {\bf X} {\bf X}^\dagger = {\bf I}$.  

Next define a $(\ell m+2)$$\times$$(\ell m+2)$ transformation matrix 
${\bf Z}_\ell$ of the form (written for $\ell$$=$$4$) 
\begin{equation}
\label{EQ:AD:S3:Z}
{\bf Z}_4 \> = \> 
\left(\begin{array}{cccccc}
1 & {\vec 0}^\dagger & {\vec 0}^\dagger & {\vec 0}^\dagger & {\vec 0}^\dagger & 0 \\
{\vec 0} & {\bf X} & {\bf 0} & {\bf 0} & {\bf 0} & {\vec 0} \\
{\vec 0} & {\bf 0} & {\bf X} &  {\bf 0} & {\bf 0} & {\vec 0} \\
{\vec 0} & {\bf 0} & {\bf 0} & {\bf X} &  {\bf 0} & {\vec 0} \\
{\vec 0} & {\bf 0} & {\bf 0} & {\bf 0} & {\bf X} & {\vec 0} \\
0 & {\vec 0}^\dagger & {\vec 0}^\dagger & {\vec 0}^\dagger & {\vec 0}^\dagger & 1 \! \\
\end{array}\right)
\end{equation}
which has the property ${\bf Z}^\dagger {\bf Z}={\bf Z} {\bf Z}^\dagger={\bf I}$.  
Multiply the equations of the form of Eq.~(\ref{EQ:AD:S2:Tlarge}) 
on the right by ${\bf Z}_\ell^\dagger$, and also 
insert the identity ${\bf I}$$=$${\bf Z}_\ell {\bf Z}_\ell^\dagger$ 
between the matrix ${\bf M}_\ell$ and the vector containing the wavefunctions 
at each site in the nanodevice.  Written for $\ell$$=$$4$, this gives
\begin{widetext}
\begin{equation}
\label{EQ:AD:S3:TlargeX}
{\bf Z}_4^\dagger {\bf M}_4 {\bf Z}_4 {\bf Z}_4^\dagger
\left(\!\begin{array}{c}
\! \delta + r \delta^* \! \\
{\vec \psi}_1 \\
{\vec \psi}_2 \\
{\vec \psi}_3 \\
{\vec \psi}_4 \\
t_T \\
\end{array}\!\right)
= 
\left(\begin{array}{cccccc}
\! \xi_w & {\vec w}^\dagger{\bf X} & {\vec 0}^\dagger & {\vec 0}^\dagger & {\vec 0}^\dagger & 0 \\
{\bf X}^\dagger{\vec w} & {\bf X}^\dagger{\bf F}_1{\bf X} & {\bf X}^\dagger{\bf B}_{12}{\bf X} & {\bf 0} & {\bf 0} & {\vec 0} \\
{\vec 0} & {\bf X}^\dagger{\bf B}_{12}^\dagger{\bf X} & {\bf X}^\dagger{\bf F}_2{\bf X} &  
                               {\bf X}^\dagger{\bf B}_{23}{\bf X} & {\bf 0} & {\vec 0} \\
{\vec 0} & {\bf 0} & {\bf X}^\dagger{\bf B}_{23}^\dagger{\bf X} & {\bf X}^\dagger{\bf F}_3{\bf X} &  
                               {\bf X}^\dagger{\bf B}_{34}{\bf X} & {\vec 0} \\
{\vec 0} & {\bf 0} & {\bf 0} & {\bf X}^\dagger{\bf B}_{34}^\dagger{\bf X} & {\bf X}^\dagger{\bf F}_4{\bf X} & {\bf X}^\dagger{\vec u} \\
0 & {\vec 0}^\dagger & {\vec 0}^\dagger & {\vec 0}^\dagger & {\vec u}^\dagger{\bf X} & \! \xi_u \! \\
\end{array}\right)
\!\!
\left(\begin{array}{c}
\! \delta + r \delta^* \! \\
{\bf X}^\dagger{\vec \psi}_1 \\
{\bf X}^\dagger{\vec \psi}_2 \\
{\bf X}^\dagger{\vec \psi}_3 \\
{\bf X}^\dagger{\vec \psi}_4 \\
t_T \\
\end{array}\right)
\! = \!
{\bf Z}_4^\dagger \!\!
\left(\begin{array}{c}
\!\! \Lambda \!\! \\
\!\! {\vec 0} \!\! \\
\!\! {\vec 0} \!\! \\
\!\!{\vec 0} \!\! \\
\!\! {\vec 0} \!\! \\
\!\! 0 \!\! \\
\end{array}\!\right)
\! = \!
\left(\begin{array}{c}
\!\! \Lambda \!\! \\
\!\! {\vec 0} \!\! \\
\!\! {\vec 0} \!\! \\
\!\!{\vec 0} \!\! \\
\!\! {\vec 0} \!\! \\
\!\! 0 \!\! \\
\end{array}\!\right)
\>.
\end{equation}
\end{widetext}
Because ${\vec x}_1$ is an eigenvector of all ${\bf B}_{j,j+1}$ 
with eigenvalue $\eta_j$ in Eq.~(\ref{EQ:S3:Beigenvalues}) 
\begin{equation}
\label{EQ:AD:S3:sMap}
{\bf X}^\dagger{\bf B}_{j,j+1}{\bf X} 
=
\left(
\begin{array}{cl}
\eta_{j} & {\vec 0}^\dagger \\
\!\!\!\!\!\!{\vec 0} & {\widetilde {\bf B}}_{j,j+1}
\end{array}
\right)
=
\left(
\begin{array}{cl}
-{\tilde s}_{j} & {\vec 0}^\dagger \\
\!\!\!\!\!\!{\vec 0} & {\widetilde {\bf B}}_{j,j+1}
\end{array}
\right)
\end{equation}
for some $(m-1)$$\times(m-1)$ matrix  ${\widetilde {\bf B}}_{j,j+1}$
which will not enter into the calculation of ${\cal T}$. 
Note we have defined ${\tilde s}_j=-\eta_j$.  
Similarly, from Eq.~(\ref{EQ:S3:Aeigenvalues}) and 
${\bf F}_j={\bf A}_j-E{\bf I}$ because ${\vec x}_1$ is an eigenvector of 
all ${\bf A}_j$, with eigenvalue $\lambda_j$, one has 
\begin{equation}
\label{EQ:AD:S3:kappaMap}
{\bf X}^\dagger{\bf F}_j{\bf X} 
=
\left(
\begin{array}{ccccc}
{\tilde\kappa}_j & {\vec 0}^\dagger \\
{\vec 0} & {\widetilde {\bf F}}_j
\end{array}
\right)
\end{equation}
with ${\tilde \kappa}_j=\lambda_j-E$ and where ${\widetilde {\bf F}}_j$
is some $(m-1)$$\times(m-1)$ matrix  
which will not enter into the calculation of ${\cal T}$. 

We also choose the connections to the nanodevice to be proportional 
to ${\vec x}_1$, so 
\begin{equation}
\label{EQ:AD:S3:wANDuMAP}
\begin{array}{lclcclc}
{\bf X}^\dagger{\vec w} 
& = & 
-t_w {\bf X} {\vec x}_1
& = & 
\left(\begin{array}{c}
-t_w \\ 0 \\ \vdots \\ 0 
\end{array}\right)
\\
{\bf X}^\dagger{\vec u} 
& = & 
-t_u {\bf X} {\vec x}_1
& = & 
\left(\begin{array}{c}
-t_u \\ 0 \\ \vdots \\ 0 
\end{array}\right)
\end{array}
\end{equation}
for some proportionality constants we label as $t_w$ and $t_u$.  

From Eq.~(\ref{EQ:AD:S2:Tlarge}) we can calculate the 
transmission ${\cal T}=|t_T|^2$ from the inverse of 
the $(\ell m +2)$$\times$$(\ell m+2)$ matrix ${\bf M}_\ell$ as 
\begin{equation}
\label{EQ:AD:S3:PRBeq10small}
\left(\begin{array}{c}
\delta+r\delta^* \\
{\vec \psi}_1 \\
\vdots \\
{\vec \psi}_\ell \\
t_T \\
\end{array}\right)
= {\bf M}^{-1}_\ell 
\left(\begin{array}{c}
\Lambda \\
{\vec 0} \\
\vdots \\
{\vec 0} \\
0 \\
\end{array}\right)
\>.
\end{equation}

However, with the transformation in Eq.~(\ref{EQ:AD:S3:TlargeX})
a large part of the matrix is decoupled from both leads, namely the 
parts which were labeled 
${\widetilde {\bf F}}_j$ and ${\widetilde {\bf B}}_{j,j+1}$.  
Therefore, we can also obtain the same transmission ${\cal T}=|t_T|^2$ 
from the inverse of the $(\ell+2)$$\times$$(\ell+2)$ matrix 
${\widetilde{\bf M}}_\ell$, which written for $\ell$$=$$4$ is 
\begin{equation}
\label{EQ:AD:S3:PRBeq10large}
\left(\begin{array}{c}
\delta+r\delta^* \\
{\tilde \psi}_1 \\
\vdots \\
{\tilde \psi}_\ell \\
t_T \\
\end{array}\right)
= {\widetilde{\bf M}}^{-1}_\ell 
\left(\begin{array}{c}
\Lambda \\
0 \\
\vdots \\
0 \\
0 \\
\end{array}\right)
\end{equation}
where 
\begin{equation}
\label{EQ:AD:S3:PRBeq9}
{\widetilde{\bf M}}_\ell 
\!\! = \!\!
\left(\begin{array}{ccccccc}
\xi_w & \!- t_w & 0 & \cdots & 0 & 0 & 0 \\
\!\! -t_w & {\tilde \kappa}_1 & -{\tilde s}_{1} &  & 0 & 0 & 0 \\
0 & -{\tilde s}_{1} & {\tilde \kappa}_2 & & 0 & 0 & 0 \\
\vdots & & & \ddots & & & \vdots \\
0 & 0 & 0 & &  {\tilde \kappa}_{\ell-1} & -{\tilde s}_{\ell-1} & 0 \\
0 & 0 & 0 & & -{\tilde s}_{\ell-1} & {\tilde \kappa}_{\ell} & \!-t_u \\
0 & 0 & 0 & \cdots & 0 & -t_u & \xi_u \\
\end{array}\right)
\>.
\end{equation}
Eq.~(\ref{EQ:AD:S3:PRBeq9}) is just the solution of a 1D chain of 
$\ell$ atoms with on site energy ${\tilde\kappa}_j+E=\lambda_j$ and 
nn (inter-slice) hopping strengths ${\tilde s}_j$.  Both the 2D system in Eq.~(\ref{EQ:AD:S3:PRBeq10small})
and the 1D system in Eq.~(\ref{EQ:AD:S3:PRBeq9}) have the same 
transmission ${\cal T}(E)$ for all energies $E$ which propagate through the 
semi-infinite leads.  We have therefore completed our exact mapping for 
the case of dimerized leads.  

In order to find a quantum dragon, we need to find TB parameters which turn 
the equivalent 1D mapped system of Eq.~(\ref{EQ:AD:S3:PRBeq9}) into 
a portion of length $\ell$ of the semi-infinite leads.  This can be accomplished by 
insisting that the original 2D device had TB parameters such that after mapping 
we have the quantum dragon condition for the connections 
\begin{equation}
\label{EQ:AD:S3:wANDuDRAGON}
t_w = t_u = t_{eo} 
\end{equation}
where $t_{eo}$ is the hopping strength in the leads from 
even-numbered to odd-numbered atoms.  
The quantum dragon conditions for the intra-slice parts of the Hamiltonian are
\begin{widetext}
\begin{equation}
\label{EQ:AD:S3:kappaDragon}
{\tilde\kappa}_j = 
\lambda_{j} - E = 
\epsilon_j-2 t_j \cos\left(\frac{\pi}{m+1}\right) - E =
\left\{
\begin{array}{lclcl}
\epsilon_o-E & \qquad\qquad & j\>{\rm odd} \\
\epsilon_e-E & \qquad\qquad & j\>{\rm even} \\
\end{array}
\right.
\end{equation}
for $j=1,2,\cdots,\ell$.  This can be satisfied by choosing the $t_j$ at random from any 
distribution (keeping $t_j$$>$$0$), and then adjusting 
\begin{equation}
\label{Eq:AD:S3:kappaDragon2}
\epsilon_j = \left\{
\begin{array}{lcl}
\epsilon_0 + 2 t_j \cos\left(\frac{\pi}{m+1}\right) & \qquad & j \> {\rm odd} \\
\epsilon_e + 2 t_j \cos\left(\frac{\pi}{m+1}\right) & \qquad & j \> {\rm even} \\
\end{array}
\right.
\end{equation}
which works since $\epsilon_j$ can be of either sign.  
Therefore, the intra-slice nn hopping strength $t_j$ can be any random value, 
provided one insists the $\epsilon_j$ satisfy Eq.~(\ref{Eq:AD:S3:kappaDragon2}).  
For the inter-slice terms of the Hamiltonian, the quantum dragon condition becomes 
\begin{equation}
\label{EQ:AD:S3:sDragon}
{\tilde s}_{j} = 
-\eta_{j} =
s_{nn,j} + 2s_{nnn,j}\cos\left(\frac{\pi}{m+1}\right)
=
\left\{
\begin{array}{lclcl}
t_{oe} & \qquad\qquad & j\>{\rm odd} \\
t_{eo} & \qquad\qquad & j\>{\rm even} \\
\end{array}
\right.
\end{equation}
\end{widetext}
for $j=1,2,\cdots,\ell-1$.  However, we want to keep the tuned values of 
$s_{nn,j}$ and $s_{nnn,j}$ both positive, and remember both $t_{oe}$ and $t_{eo}$ 
are positive.  The negative signs for the hopping terms 
have been put into the calculation by hand.  
This can be accomplished by for each $j=1,2,\cdots,\ell-1$ 
choosing two random non-negative numbers $r_{nn,j}$ and $r_{nnn,j}$, and 
setting the 2D device hopping to be 
\begin{equation}
\label{Eq:Norm2DnnOnnn}
\begin{array}{lcl}
s_{nn,j} & = & \frac{r_{nn,j} t_{oe}}{r_{nn,j}+2 r_{nnn,j}\cos\left(\frac{\pi}{m+1}\right)} \\
s_{nnn,j} & = & \frac{r_{nnn,j} t_{oe}}{r_{nn,j}+2 r_{nnn,j}\cos\left(\frac{\pi}{m+1}\right)} \\
\end{array}
\end{equation}
for $j$ odd and 
\begin{equation}
\label{Eq:Norm2DnnEnnn}
\begin{array}{lcl}
s_{nn,j} & = & \frac{r_{nn,j} t_{eo}}{r_{nn,j}+2 r_{nnn,j}\cos\left(\frac{\pi}{m+1}\right)} \\
s_{nnn,j} & = & \frac{r_{nnn,j} t_{eo}}{r_{nn,j}+2 r_{nnn,j}\cos\left(\frac{\pi}{m+1}\right)} \\
\end{array}
\end{equation}
for $j$ even.  This completes the quantum dragon conditions, which when satisfied 
gives ${\cal T}(E)$$=$$1$ for all energies which propagate through the leads.  

\begin{center}
\begin{figure}[tbh]
\vspace{0.05 cm}
\includegraphics[width=0.40\textwidth]{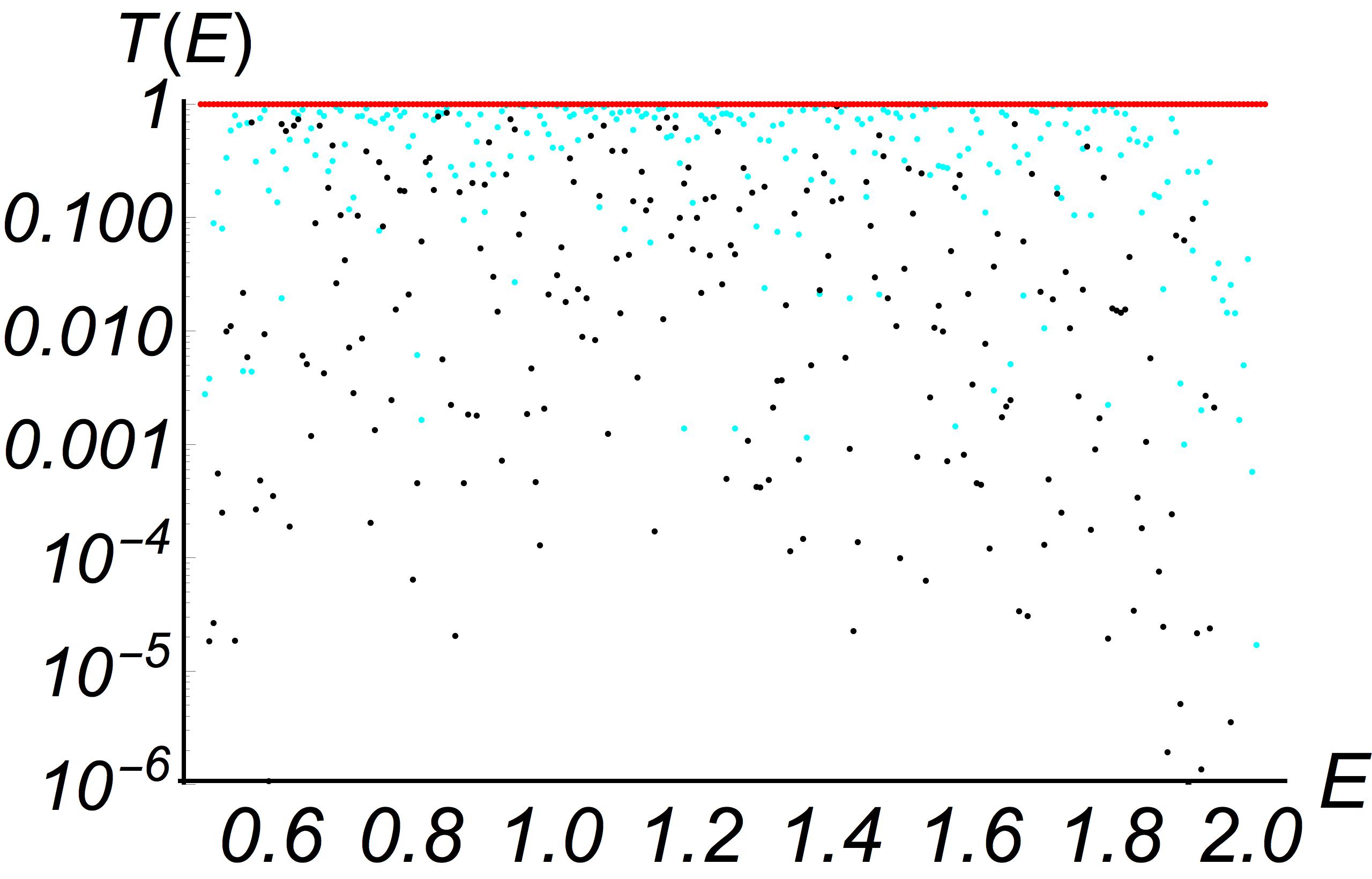}
\caption{
\label{Fig:SQdragon:Disorder:Dimerized}
(Color online.) 
An example of dimerized leads connected to a disordered, rectangular nanodevice
with $\ell$$=$$500$ and $m$$=$$7$.  
Transmission as a function of energy for the correlated disorder (red), 
showing the quantum dragon condition ${\cal T}(E)=1$.  The three 
values shown have correlated disorder so $\Delta=0$ (red dots), 
and added on~site uncorrelated disorder of strength 
$\Delta$$=$$0.05$ (cyan dots) and $\Delta$$=$$0.10$ (blue dots).  See the text for a full description.
}
\end{figure}
\end{center}

Fig.~\ref{Fig:SQdragon:Disorder:Dimerized} is an example of a quantum dragon with 
dimerized leads and a dimerized device.  The device has $\ell$$=$$500$ slices each with 
$m$$=$$7$ atoms.  The leads have $t_{eo}$$=$$0.8$ and $t_{oe}$$=$$1.2$ together with 
lead on site energies $\epsilon_o$$=$$-0.3$ and $\epsilon_e$$=$$0.3$ (remember we have 
set our zero of energy at the midpoint between the on site energies of the even and odd 
numbered leads).  
For these leads, from Eq.~(\ref{Eq:AppC:05}) and (\ref{Eq:AppC:06}), 
the leads allow electron transmission for energies in the ranges 
$-\frac{\sqrt{409}}{10}\le E\le -\frac{1}{2}$ and 
$\frac{1}{2}\le E\le \frac{\sqrt{409}}{10}\approx2.02237$. Only the positive energy 
range is shown in Fig.~\ref{Fig:SQdragon:Disorder:Dimerized}.  
Although any distribution of intra-slice hopping could be used, here 
the $t_j$ were chosen uniformly in $[0,2 t_{eo}]$ for even numbered slices and 
in $[0,2 t_{oe}]$ in odd numbered slices.  Then the on~site energy was set to the 
quantum dragon condition in Eq.~(\ref{Eq:AD:S3:kappaDragon2}).  
Similarly for the inter-slice hopping the 
$s_{nn,j}$ and $s_{nnn,j}$ were taken to be uniformly distributed in 
$[0,2 t_{eo}]$ for even $j$ and 
in $[0,2 t_{oe}]$ for odd $j$, and then tuned to the quantum dragon 
conditions as in Eq.~(\ref{Eq:Norm2DnnOnnn}) for $j$ odd 
and Eq.~(\ref{Eq:Norm2DnnEnnn}) for $j$ even.  
A total of 251 different energies uniformly spaced between 
$0.50001$ and $2.0223$ were calculated. 
Figure~\ref{Fig:SQdragon:Disorder:Dimerized} shows for the quantum dragon condition 
for the TB parameters all 
energies have ${\cal T}(E)=1$ (red dots, which overlap to look like a line segment).  
Additional uncorrelated random disorder was also included, every site having 
a different on site added disorder found by choosing a random variant from a 
normal distribution of mean zero and standard deviation unity, and then multiplying by a 
value $\Delta$.  
Figure~\ref{Fig:SQdragon:Disorder:Dimerized} also shows the, usually very small, 
values for ${\cal T}(E)$ obtained for $\Delta$$=$$0.05$ (cyan dots) and 
$\Delta$$=$$0.10$ (blue dots).  

\appendix
\section*{\label{AppE} Appendix E: Relationship between matrix and Green's function methods}
The relationship for dimerized single-channel leads, between the traditional 
Green's function method 
\cite{DATTA1995,FerryGoodnick1997,TODO2002,DATTA2005,ZIMB2011,CUEV2017,TRIO2016}
of solution and the matrix method of solution of the 
time-independent Schr{\" o}dinger equation is presented.  The 
leads have possibility for dimerized hopping ($t_{eo}$ and $t_{oe}$) and dimerized 
on~site energies ($\epsilon_e$ and $\epsilon_o$).  

The matrix method for dimerized leads, related to Eq.~(\ref{EQ:S2:Tlarge}), has the 
general form
\begin{equation}
\label{Eq:AppD:01}
\left(\begin{array}{ccc}
\xi_w & {\vec w}^\dagger & 0 \\
{\vec w} & {\cal H}-E{\bf I} & {\vec u} \\
0 & {\vec u}^\dagger & \xi_u
\end{array}\right) 
\left(\begin{array}{c}
\delta +r\delta^* \\ {\vec \psi} \\ t_T \\
\end{array}\right) 
=
\left(\begin{array}{c}
\Lambda \\ {\vec 0} \\ 0 \\
\end{array}\right) 
\end{equation}
and requires one to find the inverse of the matrix 
in order to find $t_T$.  
The transmission for any $E$ is then easily calculated by 
${\cal T}$$=$$\left|t_T\right|^2$.  
The matrix in Eq.~(\ref{Eq:AppD:01}) is not Hermitian, 
even for uniform leads due to the $\xi_w$ and $\xi_u$ 
factors.  The matrix ${\cal H}$ is the Hamiltonian of 
the nanodevice, and is therefore Hermitian.  

Any block-tridiagonal matrix of the form above 
has an inverse matrix that can easily be written as
\begin{widetext}
\begin{equation}
\label{Eq:AppD:02}
\left(\begin{array}{rrr}
\xi^{-1}_w+\xi^{-2}_w{\vec w}^\dagger{\bf L}{\vec w} & 
\>\> -\xi^{-1}_w {\vec w}^\dagger {\bf L}  & 
\xi^{-1}_w\xi^{-1}_u{\vec w}^\dagger{\bf L}{\vec u} 
\\
- \xi^{-1}_w {\bf L} {\vec w} & 
{\bf L} & 
- \xi^{-1}_u {\bf L} {\vec u} 
\\
\xi^{-1}_w \xi^{-1}_u {\vec u}^\dagger{\bf L}{\vec w} & 
- \xi^{-1}_u {\vec u}^\dagger {\bf L}  & 
\>\>
\xi^{-1}_u + \xi^{-2}_u {\vec u}^\dagger {\bf L} {\vec u}
\\
\end{array}\right) 
\left(\begin{array}{ccc}
\xi_w & {\vec w}^\dagger & 0 \\
{\vec w} & {\cal H}-E{\bf I} & {\vec u} \\
0 & {\vec u}^\dagger & \xi_u
\end{array}\right) 
\> = \> 
\left(\begin{array}{ccc}
1         & {\vec 0}^\dagger & 0 \\
{\vec 0}  &   {\bf I}        & {\vec 0} \\
0         & {\vec 0}^\dagger &  1 \\
\end{array}\right) 
\end{equation}
\end{widetext}
with the definition 
\begin{equation}
\label{Eq:AppD:03}
{\bf L} \> = \> \left[{\cal H}-E{\bf I} -
\xi^{-1}_w {\vec w}{\vec w}^\dagger - 
\xi^{-1}_u {\vec u}{\vec u}^\dagger\right]^{-1}
\>.
\end{equation}
This gives 
${\bf L}\left({\cal H}-E{\bf I}\right)=
{\bf I} + \xi^{-1}_w{\bf L}{\vec w}{\vec w}^\dagger + \xi^{-1}_u {\bf L}{\vec u}{\vec u}^\dagger$, 
which is useful in showing Eq.~(\ref{Eq:AppD:02}).  
Therefore, we have calculated the inverse of the matrix in 
Eq.~(\ref{Eq:AppD:01}), namely the matrix on the left in Eq.~(\ref{Eq:AppD:02}).

Multiplying through in Eq.~(\ref{Eq:AppD:01}) by the matrix inverse, one obtains
\begin{equation}
\label{Eq:AppD:04}
\left(\begin{array}{c}
\delta + r \delta^* \\ {\vec \psi} \\ t_T \\
\end{array}\right) 
=
\Lambda \> \xi^{-1}_w \> 
\left(\begin{array}{r}
1 + \xi^{-1}_w {\vec w}^\dagger {\bf L}{\vec w}  \\ 
- {\bf L}{\vec w} \\ 
\xi^{-1}_u {\vec u}^\dagger {\bf L} {\vec w} \\
\end{array}\right) 
\end{equation}
and consequently the transmission is
\begin{equation}
\label{Eq:AppD:05}
{\cal T}(E) \> = \> \left|t_T\right|^2 
\> = \> \left|\Lambda\right|^2 \left|\xi^{-1}_w\right|^2 \left|\xi^{-1}_u\right|^2
\left|{\vec u}^\dagger{\bf L} {\vec w}\right|^2
\>. 
\end{equation}

In the case of uniform leads the on~site energies $\epsilon_e$ and $\epsilon_o$ are set to zero, 
giving the zero of energy.  
Furthermore, for uniform leads $t_{eo}=t_{oe}=1$, setting the unit of energy.  
Then $\xi_w=\xi_u=\xi$ 
and $\xi^{-1}=\xi^*$ and $\left|\xi^{-1}\right|^2=1$, as well as 
$\Lambda=-2 i \sin(q)$, giving the transmission 
\begin{equation}
\label{Eq:AppD:06}
{\cal T}(E) \> = \> \left|t_T\right|^2 
\> = \> 4 \sin^2(q) \left|{\vec u}^\dagger{\bf L} {\vec w}\right|^2
\>. 
\end{equation}

In the Green's function formalism\cite{DATTA1995,FerryGoodnick1997,TODO2002,DATTA2005,ZIMB2011,CUEV2017,TRIO2016}, 
for the dimerized leads, 
the self-energy matrix for the incoming lead is 
\begin{equation}
\label{Eq:AppD:07}
{\bf \Sigma}_1 = -\xi^{-1}_w {\vec w}{\vec w}^\dagger
\end{equation}
and for the outgoing lead
\begin{equation}
\label{Eq:AppD:08}
{\bf \Sigma}_2 = -\xi^{-1}_u {\vec u}{\vec u}^\dagger
\end{equation}
and the Green's function is
\begin{equation}
\label{Eq:AppD:09}
\begin{array}{lcl}
{\bf{\cal G}}& \> = \> & -{\bf L} \\
& = & - \left[{\cal H}-E{\bf I} -
\xi^{-1}_w {\vec w}{\vec w}^\dagger - 
\xi^{-1}_u {\vec u}{\vec u}^\dagger\right]^{-1}
\\
& = & \left(E {\bf I} - {\cal H}  - {\bf\Sigma}_1 - {\bf\Sigma}_2 \right)^{-1}
\> .
\\
\end{array}
\end{equation}
Furthermore, since ${\vec u}^\dagger{\bf L}{\vec w}$ is a number, one can use
$\left({\vec u}^\dagger{\bf L}{\vec w}\right)^* =
\left({\vec u}^\dagger{\bf L}{\vec w}\right)^\dagger =
{\vec w}^\dagger{\bf L}^\dagger{\vec u} $
to obtain
\begin{equation}
\label{Eq:AppD:10}
\begin{array}{lcl}
\left|{\vec u}^\dagger{\bf L} {\vec w}\right|^2
& \> = \> & 
{\vec u}^\dagger{\bf L} {\vec w} \left({\vec u}^\dagger{\bf L} {\vec w}\right)^*
\\
& = & 
{\vec u}^\dagger{\bf L} {\vec w} {\vec w}^\dagger{\bf L}^\dagger {\vec u}
\\
& = & 
{\rm Tr}\left({\vec u}^\dagger{\bf L} {\vec w} {\vec w}^\dagger{\bf L}^\dagger {\vec u} \right) 
\\
& = & 
{\rm Tr}\left({\bf L} {\vec w} {\vec w}^\dagger{\bf L}^\dagger {\vec u} {\vec u}^\dagger \right) 
\\
& = & 
{\rm Tr}\left({\bf {\cal G}} {\vec w} {\vec w}^\dagger{\bf {\cal G}}^\dagger {\vec u} {\vec u}^\dagger \right) 
\\
\end{array}
\end{equation}
where ${\rm Tr}(\cdot)$ is the trace of the matrix, and the cyclic property of matrices 
within the trace have been used.  

In the normal fashion for Green's function 
calculations 
\cite{DATTA1995,FerryGoodnick1997,TODO2002,DATTA2005,ZIMB2011,CUEV2017,TRIO2016}, 
this gives for both dimerized and uniform leads
\begin{equation}
\label{Eq:AppD:11}
\begin{array}{lcl}
{\bf\Gamma}_{1}
& \> = \> & i\left({\bf\Sigma}_{1}-{\bf\Sigma}_{1}^{\dagger}\right)
\\
& = & -i \left[
\xi^{-1}_w-\left(\xi^{-1}_w\right)^*
\right] {\vec w}{\vec w}^\dagger
\\
& = & -i \left[
\frac{t_{oe}e^{i q}+ t_{eo}e^{-i q}}{t_{eo}\left(\epsilon_e-E\right) e^{- i q}}
-\frac{t_{oe}e^{-i q}+t_{eo}e^{iq}}{t_{eo}\left(\epsilon_e-E\right)e^{i q}}
\right] {\vec w}{\vec w}^\dagger
\\
& = & -i\left[
\frac{t_{oe} e^{2 i q}+t_{eo}-t_{oe}e^{-2 i q} - t_{eo}}
{t_{eo}\left(\epsilon_e-E\right)e^{i q} e^{- i q}}
\right] {\vec w}{\vec w}^\dagger
\\
& = & -i\left[
\frac{t_{oe} \left(2 i\right)\left(\frac{e^{2 i q}-e^{- 2 i q}}{2 i}\right)}{t_{eo} \left(\epsilon_e-E\right)}
\right] {\vec w}{\vec w}^\dagger 
\\
& = & \frac{2 t_{oe}}{t_{eo} \left(\epsilon_e-E\right) } \> \sin\left(2q\right) \> {\vec w}{\vec w}^\dagger
\\
& = & \gamma_1 \> {\vec w}{\vec w}^\dagger
\end{array}
\end{equation} 
and similarly
\begin{equation}
\label{Eq:AppD:12}
\begin{array}{lcl}
{\bf \Gamma}_{2}
& \> = \> & 
i \left({\bf \Sigma}_{2}-{\bf\Sigma}_{2}^{\dagger}\right) 
\\
& = & -i\left[
\xi^{-1}_u-\left(\xi^{-1}_u\right)^*
\right] \> {\vec u}{\vec u}^\dagger
\\
& = & -i\left[
\frac{t_{oe}e^{i q}+t_{eo} e^{- i q}}{t_{eo} \left(\epsilon_0-E\right) e^{-i q}}
- \frac{t_{oe} e^{- i q}+t_{eo} e^{i q}}{t_{eo}\left(\epsilon_o-E\right)e^{i q}}
\right] \> {\vec u}{\vec u}^\dagger 
\\
& = & -i\left[
\frac{t_{oe}e^{2 i q}+t_{eo}-t_{oe}e^{- 2 i q}- t_{eo}}{t_{eo}\left(\epsilon_0-E\right) e^{i q} e^{- i q}}
\right] \> {\vec u}{\vec u}^\dagger 
\\
& = & -i\left[
\frac{t_{oe}\left(2 i\right)}{t_{eo}\left(\epsilon_o-E\right)} \frac{e^{2 i q}-e^{- 2 i q}}{2 i}
\right] \> {\vec u}{\vec u}^\dagger 
\\
& = & \frac{2 t_{oe}}{t_{eo} \left(\epsilon_o-E\right)} \> \sin\left(2q\right) \> {\vec u}{\vec u}^\dagger
\\
& = & \gamma_2 \> {\vec u}{\vec u}^\dagger
\end{array}
\end{equation} 
which defines $\gamma_1$ and $\gamma_2$.  

Therefore, we only need to show to complete the equivalence between the Green's function 
and matrix methods that 
$\left|\xi^{-1}_w\right|^2 \left|\xi^{-1}_u\right|^2
\Lambda\Lambda^*=\gamma_1\gamma_2^*$.
This is shown via
\begin{widetext}
\begin{equation}
\label{Eq:AppD:13}
\begin{array}{lcl}
\left|\xi^{-1}_w\right|^2 \left|\xi^{-1}_u\right|^2 
& \> = \> &
\frac{\left(t_{oe}e^{i q}+t_{eo}e^{- i q}\right)
\left(t_{oe}e^{-i q}+t_{eo}e^{i q}\right)}
{t_{eo}^2\left(\epsilon_e-E\right)^2}
\frac{\left(t_{oe}e^{i q}+t_{eo}e^{- i q}\right)
\left(t_{oe}e^{-i q}+t_{eo}e^{i q}\right)}
{t_{eo}^2\left(\epsilon_o-E\right)^2}
\\
& = & 
\left[\frac{\left(t_{oe}e^{i q}+t_{eo}e^{- i q}\right)
\left(t_{oe}e^{-i q}+t_{eo}e^{i q}\right)}
{t_{eo}^2 \left(\epsilon_e-E\right)\left(\epsilon_o-E\right)}\right]^2
\\
& = & 
\left[\frac{t_{eo}^2+t_{oe}^2+2t_{eo}t_{oe}\cos(2q)}
{t_{eo}^2 \left(\epsilon_e-E\right)\left(\epsilon_o-E\right)}\right]^2
\\
& = & 
\left[\frac{\left(\epsilon_e-E\right)\left(\epsilon_o-E\right)}
{t_{eo}^2 \left(\epsilon_e-E\right)\left(\epsilon_o-E\right)}\right]^2
\\
& = & \frac{1}{t_{eo}^4}
\end{array}
\end{equation}
where use has been made of 
$\left(\epsilon_e-E\right)\left(\epsilon_o-E\right)
=t_{eo}^2+t_{oe}^2+2t_{eo}t_{oe}\cos(2q)$.
In addition, one has
\begin{equation}
\label{Eq:AppD:14}
\begin{array}{lcl}
\Lambda \Lambda^* & \> = \> & 
t_{oe}^2 \left(\frac{\delta}{\delta^*}e^{i q}- e^{- i q}\right) 
\left(\frac{\delta^*}{\delta}e^{-i q}- e^{i q}\right)
\\
& = & 
t_{oe}^2 \left(
2 -\frac{\delta}{\delta^*} e^{2 i q} - \frac{\delta^*}{\delta} e^{- 2 i q}
\right)
\\
& = & 
t_{oe}^2 \left[
2 \frac{\left(\epsilon_e-E\right)\left(\epsilon_o-E\right)}
{\left(\epsilon_e-E\right)\left(\epsilon_o-E\right)}
- \frac{\left(t_{eo}e^{i q}+t_{oe}e^{- i q}\right)\left(t_{oe}e^{- i q}+t_{eo} e^{i q}\right)}
{\left(\epsilon_e-E\right)\left(\epsilon_o-E\right)} e^{2 i q}
- \frac{\left(t_{eo}e^{- i q}+t_{oe}e^{iq}\right)\left(t_{oe}e^{i q}+t_{eo}e^{- i q}\right)}
{\left(\epsilon_e-E\right)\left(\epsilon_o-E\right)} e^{- 2 i q}
\right]
\\
& = & 
\frac{t_{oe}^2}{\left(\epsilon_e-E\right)\left(\epsilon_o-E\right)} \left[
2 \left(\epsilon_e-E\right)\left(\epsilon_o-E\right)
-\left(2 t_{eo}t_{oe} + 
t_{eo}^2 e^{2 i q}+t_{oe}^2 e^{- 2 i q}\right) e^{2 i q}
-\left(2 t_{eo}t_{oe}+ t_{eo}^2 e^{- 2 i q}+t_{oe}^2 e^{2 i q}\right) e^{- 2 i q}
\right]
\\
& = & 
\frac{t_{oe}^2}{\left(\epsilon_e-E\right)\left(\epsilon_o-E\right)} \left[
2 \left(\epsilon_e-E\right)\left(\epsilon_o-E\right)
-2 t_{eo}t_{oe}\left(e^{2 i q}+e^{- 2 i q}\right)
-t_{eo}^2\left(e^{4 i q}+e^{- 4 i q}\right)
- 2 t_{oe}^2
\right]
\\
& = & 
\frac{t_{oe}^2}{\left(\epsilon_e-E\right)\left(\epsilon_o-E\right)} \left[
2t_{eo}^2-2 t_{eo}^2 \cos(4q)
\right]
\\
& = & 
\frac{2 t_{oe}^2 t_{eo}^2}{\left(\epsilon_e-E\right)
\left(\epsilon_o-E\right)} \left[
1-\left(2\cos^2(2 q)-1\right)\right]
\\
& = & 
\frac{2 t_{oe}^2 t_{eo}^2}
{\left(\epsilon_e-E\right)\left(\epsilon_o-E\right)} \left[
2-2\cos^2(2q)
\right]
\\
& = & 
\frac{4 t_{oe}^2 t_{eo}^2 \> \sin^2(2 q)}
{\left(\epsilon_e-E\right)\left(\epsilon_o-E\right)} 
\\
& = & t_{eo}^4 \> \gamma_1 \gamma_2^*
\end{array}
\end{equation}
where use has been made of
\begin{equation}
\label{Eq:AppD:15}
 2 \left(\epsilon_e-E\right)\left(\epsilon_o-E\right)
 -4 t_{eo}t_{oe}\cos(2 q)
 - 2 t_{oe}^2
 =
2 t_{eo}^2
\end{equation}
\end{widetext}
and the double angle formula
\begin{equation}
\label{Eq:AppD:16}
\cos(4q)=2 \cos^2(2q)-1
\>.
\end{equation}
Consequently we have shown
\begin{equation}
\label{Eq:AppD:17}
\left|\xi^{-1}_w\right|^2 \left|\xi^{-1}_u\right|^2
\left|\Lambda \right|^2 = \gamma_1 \gamma_2^*
\>.
\end{equation}

Therefore, in the general case within the TB model, 
we find the matrix method to obtain ${\cal T}(E)$ from 
Eq.~(\ref{Eq:AppD:01}) for both dimerized leads and uniform leads 
to be equivalent to the Green's function method which has
\begin{equation}
\label{Eq:AppD:18}
{\cal T}(E)
=
{\rm Tr}\left({\bf\Gamma}_1 {\bf{\cal G}}{\bf \Gamma}_2 {\bf{\cal G}}^{\dagger}\right)
\>.
\end{equation} 

\appendix
\section*{\label{AppF} Appendix F: Quantum dragon solution for $\ell$$=$$2$}
The full solution is presented for dimerized leads connected to a simple two-slice ($\ell$$=$$2$, 
$m$$=$$1$) device.  The same equations for the transmission are 
also valid for uniform leads attached to an $\ell$$=$$2$ device.  
For simplicity, we take $\epsilon_o$$=$$\epsilon_e$$=$$0$, setting our zero of energy 
in the problem.  
This also means for both the incoming and outgoing leads $\xi_w$$=$$\xi_u$$=$$\xi$.  
The Green's function method is used, thereby requiring 
only use of $2$$\times$$2$ matrices.  The goal of this section is to show the full solution 
for the transmission ${\cal T}(E)$ for the general case, and show how it simplifies to the 
quantum dragon solution ${\cal T}(E)$$=$$1$.  

\begin{center}
\begin{figure}[tb]
\vspace{0.05 cm}
\includegraphics[width=0.40\textwidth]{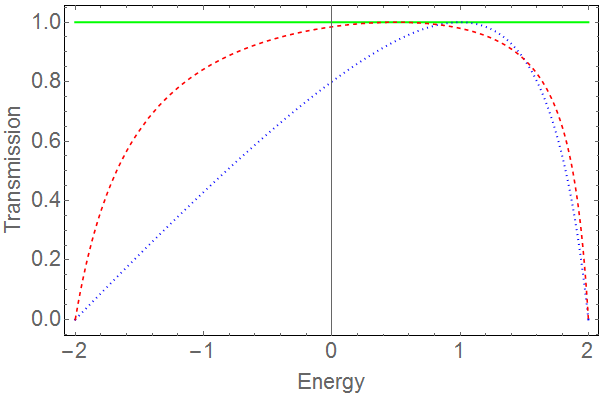}
\\
~~~
\\
\includegraphics[width=0.40\textwidth]{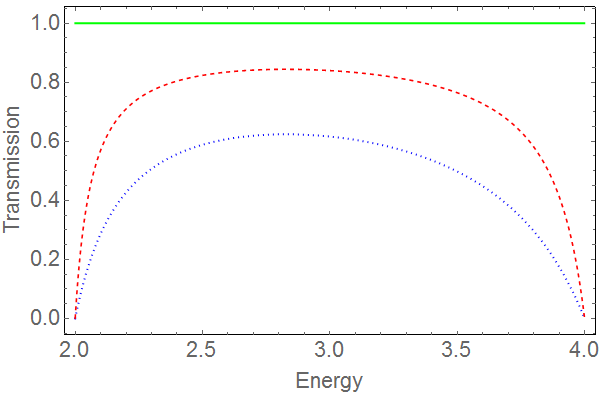}
\caption{
\label{Fig:GF_l2_m1:Fig01}
(Color online.) 
Transmission ${\cal T}$ vs energy $E$ for devices with 
$m$$=$$1$ and $\ell$$=$$2$.  
(Top) Uniform leads.  In all three curves $t_{ab}$$=$$1$.  
Shown are the three cases $\epsilon_a$$=$$\epsilon_b$$=$$1$ (blue, dotted), 
$\epsilon_a$$=$$\epsilon_b$$=$$0.5$ (red, dashed), and the quantum 
dragon solution $\epsilon_a$$=$$\epsilon_b=0$ (green, solid).  
(Bottom) Dimerized leads with $t_{eo}$$=$$1$ and $t_{oe}$$=$$3$, and  
all curves have $t_{ab}$$=$$t_{oe}$$=$$3$.  
Shown are the same three cases as in the top graph, namely 
$\epsilon_a$$=$$\epsilon_b$$=$$1.0$ (blue, dotted), 
$\epsilon_a$$=$$\epsilon_b$$=$$0.5$ (red, dashed), and the quantum 
dragon solution $\epsilon_a$$=$$\epsilon_b$$=$$0$ (green, solid) 
of ${\cal T}(E)=1$ for all energies which propagate in the 
leads.  
See text in Appendix~E for a complete description.  
}
\end{figure}
\end{center}

The Hamiltonian for the simple 2 site ($\ell$$=$$2$, $m$$=$$1$) device is
\begin{equation}
\label{Eq:AppE:01}
{\cal H}_{ab}=
\left(
\begin{array}{cc}
\epsilon_{a} & -t_{ab}\\
 -t_{ab} & \epsilon_{b} \\
\end{array}
\right)
\end{equation} 
with hopping $t_{ab}$ between the two atoms in the device, and on~site 
energies for the two atoms $\epsilon_a$ and $\epsilon_b$.  
The device is coupled to two dimerized single-channel leads [incoming 
vector ${\vec w}^\dagger=\left( -w \> 0 \right)$ and outgoing 
vector ${\vec u}^\dagger=\left( 0 \> -\! u \right)$], 
giving the self energy matrices  
\begin{equation}
\label{Eq:AppE:02}
\Sigma_{1}=
\left(
\begin{array}{cc}
\frac{-w^{2}}{\xi} & 0 \\
0 & 0 \\
\end{array}
\right)
\qquad 
{\rm and} \qquad
\Sigma_{2}=
\left(
\begin{array}{cc}
0 & 0 \\
0 & \frac{-u^{2}}{\xi} \\
\end{array}
\right)
\>.
\end{equation} 
The figure setup is the same as in App.~~A of ref.~\onlinecite{MANdragon2014}.  
It should be noted that the quantity $\xi$ is complex, but both $w$ and $u$ are 
here taken to be real positive numbers. 
The complex quantity $\xi$ is different for uniform leads and dimerized leads,
namely
\begin{equation}
\label{Eq:AppE:03}
\xi \> = \> \left\{ 
\begin{array}{lcl}
e^{-i q} & \qquad & {\rm uniform} \\
\frac{-E \> t_{eo} \> e^{- i q} }{ t_{eo} e^{- i q} + t_{oe} e^{iq} } & & {\rm dimerized} 
\>. \\
\end{array}
\right.
\end{equation}
From Eq.(\ref{Eq:AppE:02}), the coupling matrices are expressed as 
\begin{equation}
\label{Eq:AppE:04}
{\bf\Gamma}_{1}=i \left({\bf\Sigma}_{1}-{\bf\Sigma}_{1}^{\dagger}\right)
=iw^{2}
\left(
\begin{array}{cc}
\frac{1}{\xi^{*}}-\frac{1}{\xi} & 0 \\
0 & 0 \\
\end{array}
\right)
\end{equation} 
and 
\begin{equation}
\label{Eq:AppE:05}
{\bf \Gamma}_{2}
=
i\left({\bf \Sigma}_{2}-{\bf\Sigma}_{2}^{\dagger}\right)
=i u^{2}
\left(
\begin{array}{cc}
0 & 0 \\
0 & \frac{1}{\xi^{*}}-\frac{1}{\xi} \\
\end{array}
\right)
\>.
\end{equation} 
The Green's function, ${\bf{\cal G}}$ is 
\begin{equation}
\label{Eq:AppE:06}
\begin{array}{lcl}
{\bf{\cal G}} & = &
\left(EI-{\cal H}-{\bf\Sigma}_{1}-{\bf\Sigma}_{2}\right)^{-1} \\
& = & 
\left(
\begin{array}{cc}
\frac{w^{2}}{\xi}-\kappa_{a} & t_{ab} \\
t_{ab} & \frac{u^{2}}{\xi}-\kappa_{b}\\
\end{array}
\right)^{-1} 
\end{array}
\end{equation}
with $\kappa_a$$=$$\epsilon_a$$-$$E$ and $\kappa_b$$=$$\epsilon_b$$-$$E$.  
The electron transmission probability can be expressed in terms of the Green's function, 
${\cal G}$ and the coupling matrices, ${\bf\Gamma}_{1}$ and ${\bf\Gamma}_{2}$ as 
\begin{equation}
\label{Eq:AppE:07}
{\cal T}(E)
=
{\rm Tr}\left({\bf\Gamma}_1 {\cal G}{\bf\Gamma}_2 {\cal G}^{\dagger}\right)
\>.
\end{equation} 
Put Eq.~(\ref{Eq:AppE:04}), Eq.~(\ref{Eq:AppE:05}) and 
Eq.~(\ref{Eq:AppE:06}) into Eq.(\ref{Eq:AppE:07}). 
This gives the electron transmission probability as 
\begin{widetext}
\begin{equation}
\label{Eq:AppE:08}
{\cal T}(E)=
\frac{ w^{2} u^{2} t_{ab}^{2} \Lambda \Lambda^* }
{\left[(w^{2}-\kappa_{a}\xi)(u^{2}-\kappa_{b}\xi)-t_{ab}^{2}\xi^{2}\right]
\left[(w^{2}-\kappa_{a}\xi^*)(u^{2}-\kappa_{b}\xi^*)-t_{ab}^{2}\xi^{*2}\right]}
\>.
\end{equation} 
In the case when $w=u$, Eq.(\ref{Eq:AppE:08}) becomes  
\begin{equation}
\label{Eq:AppE:09}
{\cal T}(E)= 
\frac{ w^{4} t_{ab}^{2} \Lambda \Lambda^* }
{\left[(w^{2}-\kappa_{a}\xi)(w^{2}-\kappa_{b}\xi)-t_{ab}^{2}\xi^{2}\right]
\left[(w^{2}-\kappa_{a}\xi^*)(w^{2}-\kappa_{b}\xi^*)-t_{ab}^{2}\xi^{*2}\right]} 
\> .
\end{equation} 
Eq.~(\ref{Eq:AppE:09}) is the transmission probability for the two site device coupled to 
either uniform ($t_{eo}$$=$$t_{oe}$$=$$t_{lead}$) or dimerized ($t_{eo}$$\ne$$t_{oe}$) 
leads, depending on the value of 
$\xi$ in Eq.~(\ref{Eq:AppE:03})
and $\Lambda$ in Eq.~(\ref{Eq:AppC:10}).  

Furthermore, when $\epsilon_a$$=$$\epsilon_b$, Eq.~(\ref{Eq:AppE:09}) becomes 
\begin{equation}
\label{Eq:AppE:10}
{\cal T}(E)= 
\frac{ w^{4} t_{ab}^{2} \Lambda \Lambda^* }
{\left[(w^{2}-\kappa_{a}\xi)^2-t_{ab}^{2}\xi^{2}\right]
\left[(w^{2}-\kappa_{a}\xi^*)^2-t_{ab}^{2}\xi^{*2}\right]} 
\>.
\end{equation} 
A plot of ${\cal T}(E)$ 
for $\ell$$=$$2$ for selected parameters is shown in Fig.~\ref{Fig:GF_l2_m1:Fig01} for both 
uniform and for dimerized leads.  

To see how the quantum dragon solution is obtained from the mathematics, consider a 
$\ell$$=$$2$ and $m=1$ device with uniform leads ($t_{oe}$$=$$t_{eo}$$=$$1$), with 
$w$$=$$u$$=$$t_{ab}$$=$$1$ and 
$\epsilon_a$$=$$\epsilon_b$$=$$0$ so $\kappa_a$$=$$\kappa_b$$=$$-E$.  
Eq.~(\ref{Eq:AppE:10}) gives the quantum dragon solution from the operations 
\begin{equation}
\label{Eq:AppE:11}
\begin{array}{lclcl}
{\cal T}(E)
& \> = \> & 
\frac{\Lambda \Lambda^* }
{\left[(1+E\xi)^{2}-\xi^{2}\right]
\left[(1+E\xi^*)^{2}-\xi^{*2}\right]} & & {\rm starting,} \> \kappa_a=\kappa_b=-E 
\\
& = & \frac{4 \sin^2(q)}{\left[(1+E e^{i q})^{2}-e^{2 i q}\right]
\left[(1+E e^{- i q})^{2}-e^{-2 i q}\right]} 
& \qquad\qquad & 
{\rm use } \> \Lambda=-2 i \sin(q) \> {\rm and} \> \xi=e^{i q} 
\\
& = & \frac{4-E^2}
{\left[1+2 E e^{i q} + \left(E^2-1\right)e^{2 i q}\right]\left[1+2 E e^{-i q} + \left(E^2-1\right)e^{-2 i q}\right]}
& & {\rm use } \> \sin(q)=\frac{\sqrt{4-E^2}}{2}, \> {\rm expand}
\\
& = & \frac{4-E^2}
{2 + 2E^2 + E^4+4 E \cos(q) +2\left(E^2-1\right)\cos(2 q)+ 4 E\left(E^2-1\right)\cos(q)}
& & {\rm multiply \> out, \> group \> terms, \> use} \> 2\cos(q)=e^{i q}+e^{-i q}
\\
& = & \frac{4-E^2}
{2 + 2E^2 + E^4 -2 E^2 +2\left(E^2-1\right)\left(\frac{E^2}{2}-1\right) - 2 E^2 \left(E^2-1\right)}
& & {\rm use} \> \cos(q)=-\frac{E}{2}, \> \cos(2q)=2\cos^2(q)-1
\\
& = & \frac{4-E^2}
{2 + 2E^2 + E^4 -2 E^2 +\left(E^4-3E^2+2\right) - 2 E^4 + 2 E^2 }
& & {\rm multiply \> out} 
\\
& = & \frac{4-E^2}{4-E^2} & & {\rm collect\> terms \> (many \> terms \> cancel)}
\\
& = & 1 \> . & & {\rm quantum \> dragon\> solution}
\end{array}
\end{equation} 
\end{widetext}
This $m=1$ and $\ell=2$ device acts as a \lq short circuit' between the 
two uniform leads.  
 Because it is a 
\lq short circuit', physically the solution ${\cal T}(E)=1$ makes 
sense.  This physical solution should also extend to the 
case of $m=1$ and general $\ell$, but the algebra becomes 
more messy than Eq.~(\ref{Eq:AppE:11}).  

Starting from Eq.~(\ref{Eq:AppE:10}) for the dimerized case, a \lq short circuit' should 
be found for even $\ell$ when $w=u=t_{eo}$, $t_{ab}=t_{oe}$, and 
$\epsilon_a=\epsilon_b=0$.  However, the algebra becomes very messy 
in this case, even for a $m=1$ and $\ell=2$ device.  
However, physically one expects a \lq short circuit' solution 
in the quantum sense, because the inserted device 
has the same structure as the leads. This is indeed what we observe numerically, 
as seen in Fig.~\ref{Fig:GF_l2_m1:Fig01}.   



\begin{thebibliography}{0}%
\makeatletter
\providecommand \@ifxundefined [1]{%
 \@ifx{#1\undefined}
}%
\providecommand \@ifnum [1]{%
 \ifnum #1\expandafter \@firstoftwo
 \else \expandafter \@secondoftwo
 \fi
}%
\providecommand \@ifx [1]{%
 \ifx #1\expandafter \@firstoftwo
 \else \expandafter \@secondoftwo
 \fi
}%
\providecommand \natexlab [1]{#1}%
\providecommand \enquote  [1]{``#1''}%
\providecommand \bibnamefont  [1]{#1}%
\providecommand \bibfnamefont [1]{#1}%
\providecommand \citenamefont [1]{#1}%
\providecommand \href@noop [0]{\@secondoftwo}%
\providecommand \href [0]{\begingroup \@sanitize@url \@href}%
\providecommand \@href[1]{\@@startlink{#1}\@@href}%
\providecommand \@@href[1]{\endgroup#1\@@endlink}%
\providecommand \@sanitize@url [0]{\catcode `\\12\catcode `\$12\catcode
  `\&12\catcode `\#12\catcode `\^12\catcode `\_12\catcode `\%12\relax}%
\providecommand \@@startlink[1]{}%
\providecommand \@@endlink[0]{}%
\providecommand \url  [0]{\begingroup\@sanitize@url \@url }%
\providecommand \@url [1]{\endgroup\@href {#1}{\urlprefix }}%
\providecommand \urlprefix  [0]{URL }%
\providecommand \Eprint [0]{\href }%
\providecommand \doibase [0]{http://dx.doi.org/}%
\providecommand \selectlanguage [0]{\@gobble}%
\providecommand \bibinfo  [0]{\@secondoftwo}%
\providecommand \bibfield  [0]{\@secondoftwo}%
\providecommand \translation [1]{[#1]}%
\providecommand \BibitemOpen [0]{}%
\providecommand \bibitemStop [0]{}%
\providecommand \bibitemNoStop [0]{.\EOS\space}%
\providecommand \EOS [0]{\spacefactor3000\relax}%
\providecommand \BibitemShut  [1]{\csname bibitem#1\endcsname}%
\let\auto@bib@innerbib\@empty
\end{thebibliography}%


\begin{thebibliography}{10}

\bibitem{WALD2016} M.M.\ Waldrop, 
{\it The chips are down for Moore's law\/}, 
Nature {\bf 530}, 141-147 (2016).  

\bibitem{TAKA2010} T.\ Tsurumi, H.\ Hirayama, M.\ Vacha, and T.\ Taniyama, 
{\it Nanoscale physics for materials science\/}
(CRC Press, Boca Raton, FL, 2010).  

\bibitem{PUER2017} 
{\it Nanoelectronics: Materials, Devices, Applications}, two volumes, 
Edited by R.\ Puers, L.\ Baldi, M.\ {v}an~{d}e~{V}oorde, and S.E.\ {v}an~{N}ooten,
(Wiley-VCH, Weinheim, Germany, 2017).  

\bibitem{DATTA1995}
Datta, S.\ 
{\it Electronic Transport in Mesoscopic Systems\/}
(Cambridge University Press, Cambridge, UK, 1995).

\bibitem{FerryGoodnick1997}
Ferry, D.K.,\  \& Goodnick, S.M.\ 
{\it Transport in Nanostructures\/}
(Cambridge University Press, Cambridge, UK, 1997).

\bibitem{TODO2002} T.N.\ Todorov,
{\it Tight-binding simulation of current-carrying nanostructures}
J.\ Phys.: Condens.\ Matter {\bf 14}, 3049-3084 (2002).

\bibitem{DATTA2005}
Datta, S.\ 
{\it Quantum Transport: Atom to Transistor\/}
(Cambridge University Press, Cambridge, UK, 2005).
  
\bibitem{ZIMB2011}
N.A.\ Zimbovskaya and M.R.\ Pederson,
{\it Electron transport through molecular junctions\/},
Phys.\ Reports {\bf 509}, 1-87 (2011).  

\bibitem{CUEV2017}
J.C.\ Cuevas and E.\ Scheer,  
{\it Molecular Electronics: An introduction to theory and experiment\/}, 
$2^{\rm nd}$ edition, 
(World Scientific, Singapore, 2017).  

\bibitem{LAND57} 
Landauer, R.\ 
{\it Spatial variation of currents and fields due to localized scatterers in 
metallic conduction.}  
{\it IBM J.\ Research and Development\/} {\bf 1}, 223-231 (1957).

\bibitem{BAGW1989}
P.F.\ Bagwell and T.P.\ Orlando,
{\it Landauer's conductance formula and its generalization to finite voltages.}
{\it Phys.\ Rev.\ B} {\bf 40}, 1456-1464 (1989).  

\bibitem{LESO1989} G.B.\ Lesovik, 
{\it Excess quantum noise in 2D ballistic point contacts\/}, 
JETP Lett.\ {\bf 49}, 592-594 (1989).

\bibitem{BUTT1990} M.\ B{\"u}ttiker, 
{\it Scattering theory of thermal and excess noise in open conductors\/}, 
Phys.\ Rev.\ Lett.\ {\bf 65}, 2901-2904 (1990).  

\bibitem{KUMA1996} 
A.\ Kumar, L.\ Saminadayar, D.C.\ Glatti, Y.\ Jin, and B.\ Etienne,
{\it Experimental test of the quantum shot noise reduction theory},
Phys.\ Rev.\ Lett.\ {\bf 76}, 2778-2781 (1996).  

\bibitem{OUIS2013}
T.\ Ouisse, {\it Electron Transport in Nanostructures and Mesoscopic Devices: An Introduction}
(John Wiley \& Sons, Hoboken, NJ, 2013).  

\bibitem{DEPI2001} 
R.\ de~{P}icciotto, H.L.\ Stormer, L.N.\ Pfeiffer, K.W.\ Baldwin, and K.W.\ West, 
{\it Four-terminal resistance of a ballistic quantum wire\/},
Nature {\bf 411}, 51-54 (2001).  

\bibitem{WAKA2007} K.\ Wakabayashi, Y.\ Takane, and M.\ Sigrist, 
{\it Perfectly conducting channel and universality crossover in disordered graphene 
nanoribbons}, Phys.\ Rev.\ Lett.\ {\bf 99}, 036601 (2007).  

\bibitem{MATS2015} A.\ Matsumoto, T.\ Arita, Y.\ Takane, 
Y.\ Yoshimura, and K.-I.\ Imura, 
{\it Manipulating quantum channels in weak topological insulator nanoarchitectures\/}, 
Phys.\ Rev.\ B {\bf 92}, 195424 [14~pages] (2015).  

\bibitem{MANdragon2014}
M.A.\ Novotny, {\it Energy-independent total quantum transmission of 
electrons through nanodevices with correlated disorder}, 
Phys.\ Rev.\ B {\bf 90}, 165103 [14 pages] (2014).

\bibitem{OUYA2001} M.\ Ouyang, J.-L.\ Huang, C.L.\ Cheung, and C.M.\ Lieber, 
{\it Energy gaps in \lq\lq metallic" single-walled carbon nanotubes},
Science {\bf 292}, 702-702 (2001).  

\bibitem{KONG2001}
J.\ Kong, E.\ Yenilmez, T.W.\ Tombler, W.\ Kim, and H.\ Dai,
{\it Quantum interference and ballistic transmission in nanotube electron waveguides},
Phys.\ Rev.\ Lett.\ {\bf 87}, 106801 [4~pages] (2001).    

\bibitem{BARI2014} J. Baringhaus, M.\ Ruan, F.\ Edler, A.\ Tejeda, M.\ Sicot, A.\ Taleb-{I}brahim,
A.-P.\ Li, Z.\ Jiang, E.H.\ Conrad, C.\ Berger, C.\ Tegenkamp, and W.A.\ de~{H}eer, 
{\it Exceptional ballistic transport in epitaxial graphene nanoribbons}, 
Nature {\bf 506}, 349-354 (2014).  

\bibitem{CELI2016} A.\ Celis, M.N.\ Nair, A.\ Taleb-Ibrahim, 
E.H.\ Conrad, C.\ Berger, W.A.~de~{H}eer, and A.\ Tejeda, 
{\it Graphene nanoribbons: fabrication properties and devices},
J.\ Phys.\ D: App.\ Phys.\ {\bf 49}, 143001 (2016).  

\bibitem{WHIT1998} C.T.\ White and T.N.\ Todorov,
{\it Carbon nanotubes as long ballistic conductors\/}, 
Nature {\bf 393}, 240-242 (1989).

\bibitem{Erts2000}
D.\ Erts, H.\ Olin, L.\ Ryen, E.\ Olsson, 
and A.\ Th{\"o}l{\'e}n, 
{\it Maxwell and {S}harvin conductance in gold point contacts investigated using {T}{E}{M}-{S}{T}{M}}
Phys.\ Rev.\ B {\bf 61}, 112725-12727 (2000).  

\bibitem{DCA2000}
Daboul, D.,\ Chang, I.,\ \& Aharony, A.\ 
Series expansion study of quantum percolation on the square lattice.
{\it Euro. J.\ Phys.\ B\/} {\bf 16}, 303-316 (2000).

\bibitem{ZHAO2014} J.\ Zhao, Q.\ Deng, A.\ Bachmatiuk, G.\ Sandeep, A.\ Popov, and J.\ Eckert, 
{\it Free-standing single-atom thick iron membranes suspended in graphene pores\/}, 
Science {\bf 343}, 1228-1232 (2014).  

\bibitem{YIN2016} K.\ Yin, Y.-Y.\ Zhang, Y.\ Zhou, L.\ Sun, M.F.\ Chisholm, 
S.T.\ Pantelides, and W.\ Zhou,
{\it Unsupported single-atom-thick copper oxide monolayers\/}, 
2D Mater.\ {\bf 4}, 011001 [8~pages] (2017).  

\bibitem{KANO2017}
E.\ Kano, D.G.\ Kvashnin, S.\ Sakai, L.A.\ Chernozatonskii, 
P.B.\ Sorokin, A.\ Hashimoto, and M.\ Takeguchi,
{\it One-atom-thick 2D copper oxide clusters on graphene\/}, 
Nanoscale {\bf 9}, 3980-3985 (2017).  

\bibitem{NOVO2016} K.S.\ Novoselov, A.\ Mishchenko, A.\ Carvalho, and A.H.\ Castro {N}eto,
{\it 2D materials and van der {W}aals heterostructures},
Science {\bf 353}, aac9439 [11~pages] (2016).  

\bibitem{BOET2011}
S.\ Boettcher, C.\ Varghese, and M.A.\ Novotny,
{\it Quantum transport through hierarchical structures},
Phys.\ Rev.\ E {\bf 83}, 041106 [12~pages] (2011).  

\bibitem{NOVO2014} 
M.A.\ Novotny, L.\ Solomon, and G.\ Inkoom,
{\it Quantum transport through a fully connected network with disorder\/}, 
Phys.\ Proceedia {\bf 53}, 71-74 (2014).  

\bibitem{TRIO2016}
{\it Simulation of transport in nanodevices\/}, 
Editors F.\ Triozon and P.\ Dollfus,
(John Wiley \& Sons, Hoboken, NJ, 2016).  

\bibitem{JAIN1979} A.K.\ Jain, 
{\it A sinusoidal family of unitary transforms\/}, 
IEEE Trans.\ Pattern Analysis and Machine Intelligence, 
{\bf PAMI-1}, 356-365 (1979).

\bibitem{ANDE1958} P.W.\ Anderson,
{\it Absence of diffusion in certain random lattices\/}, 
Phys.\ Rev.\ {\bf 109}, 1492-1505 (1958).  

\bibitem{SCHW2007}
T.\ Schwartz, G.\ Bartal, S.\ Fishman, and M.\ Segev, 
{\it Transport and {A}nderson localization in disordered two-dimensional photonic lattices},
Nature {\bf 446}, 52-55 (2007).

\bibitem{INKO2017} G.\ Inkoom, 
{\it Quantum dragon solutions for electron transport through single-layer planar 
rectangular crystals\/}, Ph.D.\ Dissertation, Mississippi State University, 2017.  

\bibitem{Hanoi2011}
S.\ Boettcher, C.\ Varghese, and M.A.\ Novotny
{\it Quantum transport through hierarchical structures}
Phys.\ Rev.\ B {\bf 83}, 041106 [12 pages] (2011).  

\bibitem{SOUM2002}
S.\ Souma and A.\ Suzuki,
{\it Local density of states and scattering matrix in quasi-one-dimensional systems\/},
Phys.\ Rev.\ B {\bf 65}, 115307 [7~pages] (2002).  

\bibitem{MIRO2010}
A.E.\ Miroshnichenko, S.\ Flach, and Y.S.\ Kivshar, 
{\it Fano resonances in nanoscale structures\/}, 
Rev.\ Mod.\ Phys.\ {\bf 82}, 2257-2298 (2010).  

\bibitem{TARA2016} V.E. Tarasov, 
{\it Exact discretization of Schr{\"o}dinger equation}, 
Physical Letters A {\bf 380}, 68-75 (2016).  

\bibitem{FLIN2008}
C.\ Flindt, T.\ Novotny{\' y}, A.\ Braggio, M.\ Sassetti, and A.-P.\ Jauho, 
{\it Counting statistics of non-{M}arkovian quantum stochastic processes},
Phys\ Rev.\ Lett.\ {\bf 100}, 150601 [4~pages] (2008).  

\bibitem{BIBO2016} 
A.\ Biborski, A.P.\ Kadzzielawa, A.\ Gorzyca-Goraj, E.\ Zipper, M.\ Ma{\'s}ka, and J.\ Spalek, 
{\it Dot-ring nanostructure: Rigorous analysis of many-electron effects\/}, 
Scientific Reports {\bf 6}, 29887 [14~pages] (2016).  

\bibitem{MANpatent} M.A.\ Novotny, 
{\it Materials and devices that provide total transmission of electrons without ballistic propagation and 
methods for devising same}, U.S.\ Patent Pending.  

\bibitem{Koni2008}
Y Konishi, H.\ Tokoro, M.\ Nishino, 
and S.\ Miyashita, 
{\it Monte Carlo Simulation of Pressure-Induced Phase 
Transitions in Spin-Crossover Materials\/}
Phys.\ Rev.\ Lett.\ {\bf 100}, 067206 [4~pages] (2008).

\bibitem{Bloch}
F.\ Bloch, 
{\it {\"U}ber die {Q}uantenmechanik der {E}lectronen in {K}ristallgittern\/}, 
Z.\ Physik {\bf 52}, 555-600 (1928).  



\end{thebibliography}
\end{document}